\PassOptionsToPackage{table, dvipsnames}{xcolor}
\documentclass[sigconf]{acmart}
\AtBeginDocument{%
  }

\usepackage{booktabs}
\usepackage{array}
\usepackage{balance} 
\usepackage{multirow}
\usepackage[normalem]{ulem}
\usepackage{color}
\definecolor{lightgray}{RGB}{215,215,215}
\usepackage{colortbl}
\useunder{\uline}{\myul}{}
\usepackage{subfigure}
\usepackage{algorithm}  
\usepackage{algorithmicx}  
\usepackage[noend]{algpseudocode}  
\usepackage{stackengine}

\usepackage{amsmath}  
\usepackage{enumitem}
\usepackage{tabularx}
\usepackage[utf8]{inputenc}
\usepackage[english]{babel}
\usepackage{amsthm}
\usepackage{bm}
\usepackage{threeparttable}

\usepackage[most]{tcolorbox}
\usepackage{graphicx}

\newcommand{\ie}{\emph{i.e., }}
\newcommand{\eg}{\emph{e.g., }}

\newcommand{\cf}{\emph{cf. }}

\newcommand\xrowht[2][0]{\addstackgap[.5\dimexpr#2\relax]{\vphantom{#1}}}

\newlength\myindent
\usepackage{balance}

\usepackage{soul}

\newcommand{\mathcolorbox}[2]{%
    \definecolor{tempcolor}{HTML}{#1}
    \colorbox{tempcolor}{$\displaystyle #2$}%
}

\floatname{algorithm}{Algorithm}

\copyrightyear{2025}
\acmYear{2025}
\setcopyright{acmlicensed}\acmConference[WWW '25]{Proceedings of the ACM Web Conference 2025}{April 28-May 2, 2025}{Sydney, NSW, Australia}
\acmBooktitle{Proceedings of the ACM Web Conference 2025 (WWW '25), April 28-May 2, 2025, Sydney, NSW, Australia}
\acmDOI{10.1145/3696410.3714843}
\acmISBN{979-8-4007-1274-6/25/04}

\begin{document}

\title{Personalized Image Generation with Large Multimodal Models}

\author{Yiyan Xu}
\email{yiyanxu24@gmail.com}
\affiliation{
\institution{University of Science and Technology of China}
\country{China}
\city{Hefei}
}

\author{Wenjie Wang}
\email{wenjiewang96@gmail.com}
\affiliation{
\institution{University of Science and Technology of China}
\country{China}
\city{Hefei}
}


\author{Yang Zhang}
\email{zyang1580@gmail.com}
\affiliation{
\institution{National University of Singapore}
\country{Singapore}
\city{Singapore}
}

\author{Biao Tang}
\email{biao.tang@meituan.com}
\affiliation{
\institution{Meituan}
\country{China}
\city{Shanghai}
}

\author{Peng Yan}
\email{yanpeng04@meituan.com}
\affiliation{
\institution{Meituan}
\country{China}
\city{Beijing}
}

\author{Fuli Feng}
\authornote{Corresponding authors. This work is supported by the National Natural Science Foundation of China (62272437, U24B20180, 62121002) and the advanced computing resources provided by the Supercomputing Center of the USTC.}
\email{fulifeng93@gmail.com}
\affiliation{
\institution{University of Science and Technology of China}
\country{China}
\city{Hefei}
}

\author{Xiangnan He}
\authornotemark[1]
\email{xiangnanhe@gmail.com}
\affiliation{
\institution{MoE Key Lab of BIPC, University of Science and Technology of China}
\country{China}
\city{Hefei}
}

\renewcommand{\shortauthors}{Yiyan Xu et al.}

\begin{abstract}
Personalized content filtering, such as recommender systems, has become a critical infrastructure to alleviate information overload. However, these systems merely filter existing content and are constrained by its limited diversity, making it difficult to meet users' varied content needs. 
To address this limitation, personalized content generation has emerged as a promising direction with broad applications. 
Nevertheless, most existing research focuses on personalized text generation, with relatively little attention given to personalized image generation. The limited work in personalized image generation faces challenges in accurately capturing users' visual preferences and needs from noisy user-interacted images and complex multimodal instructions. Worse still, there is a lack of supervised data for training personalized image generation models.  

To overcome the challenges, we propose a \textit{\textbf{P}ersonalized \textbf{I}mage \textbf{Ge}nerati\textbf{on} Framework} named Pigeon, which adopts exceptional large multimodal models with three dedicated modules to capture users' visual preferences and needs from noisy user history and multimodal instructions. 
To alleviate the data scarcity, we introduce a two-stage preference alignment scheme, comprising masked preference reconstruction and pairwise preference alignment, to align Pigeon with the personalized image generation task. 
We apply Pigeon to personalized sticker and movie poster generation, where extensive quantitative results and human evaluation highlight its superiority over various generative baselines. 

\end{abstract}



\maketitle

\vspace{-0.1cm}
\section{Introduction}
\label{sec:introduction}

\begin{figure}[t]
\setlength{\abovecaptionskip}{0.05cm}
\setlength{\belowcaptionskip}{-0.6cm}
\centering
\includegraphics[scale=0.53]{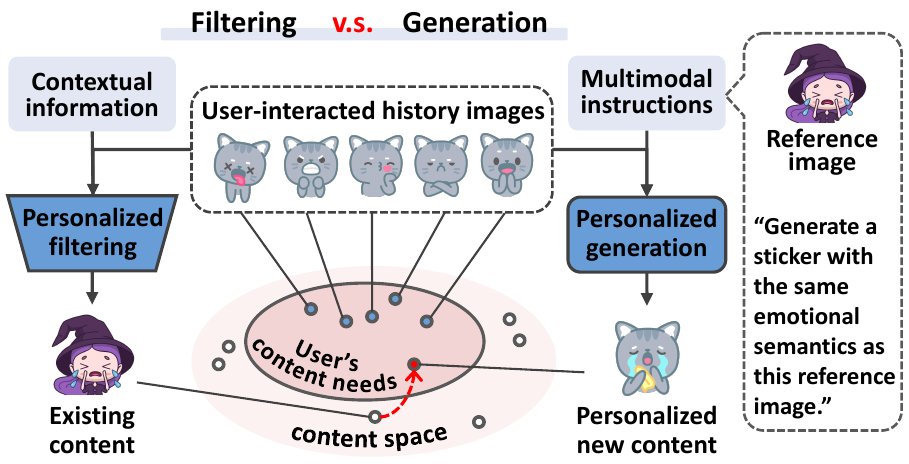}
\caption{Personalized filtering selects the most relevant existing content while personalized generation creates new and customized ones, more precisely satisfying users' diverse content needs.} 
\label{fig:intro}
\end{figure}
 
In the era of information overload, individuals are overwhelmed with vast amounts of multimodal content on the Web, underscoring the importance of personalized content delivery~\cite{wu2023personalized,wu2013data,zhu2024understanding}. 
The predominant approach, personalized content filtering like recommender systems~\cite{wang2023diffusion,he2020lightgcn,wang2019neural,gao2023cirs}, relies on user interaction history and contextual information to infer user preferences and filter existing content. 
However, this approach is constrained by the limited diversity of available content, rendering it inadequate to fully meet users' varied content needs (see an example in Figure~\ref{fig:intro}). 
To address this limitation, generating personalized new content is becoming increasingly important across various domains, including personalized movie posters~\cite{pmg}, advertisements~\cite{yang2024new,vashishtha2024chaining}, music~\cite{dai2022personalised,manos2024music}, and fashion designs~\cite{xu2024diffusion,yu2019personalized}. 

Previous works on personalized content generation primarily focus on personalized text generation~\cite{salemi-etal-2024-lamp,kumar2024longlamp,salemi2024optimization,rame2024rewarded} while personalized image generation receives little attention. 
Technically, personalized image generation aims to capture implicit user preferences from user-interacted history images and then integrate users' explicit needs from multimodal instructions to generate personalized target images, as illustrated in Figure~\ref{fig:intro}. 
Existing methods mainly rely on Diffusion Models (DMs) or Large Language Models (LLMs) for personalized image generation:  
\begin{itemize}[leftmargin=*]
    \item \textbf{DM-based methods}~\cite{textualinversion,ruiz2023dreambooth,xu2024diffusion,yang2024new,czapp2024dynamic} might learn the representations of implicit user preferences from user-interacted history images and combine these representations with explicit user instructions for target images to guide the generation of DMs. 
    However, these methods struggle to accurately capture user preferences from noisy history images, which typically cover diverse and complex user interests. 

    \item \textbf{LLM-based Personalized Multimodal Generation (PMG)}~\cite{pmg} converts history images and multimodal instructions into textual descriptions, and then utilizes pre-trained LLMs to encode textual descriptions for guiding image generation. However, the discrete nature of text makes it difficult to convey complex visual information in history images and instructions, leading to imprecise representations.
\end{itemize}

In this light, the key to personalized image generation lies in accurately inferring implicit user preferences from noisy history images while adhering to explicit multimodal instructions for image generation. 
This necessitates robust multimodal understanding, reasoning, and instruction-following capabilities, driving the adoption of Large Multimodal Models (LMMs)~\cite{lavit,ge2024making} for personalized image generation. 
An intuitive approach is to transform history images and multimodal instructions into visual and textual tokens as the input of LMMs for cross-modal understanding and image generation. 
However, this approach faces critical challenges: 
\begin{itemize}[leftmargin=*]
    \item User-preferred and disliked features (\eg characters and colors) are often entangled within user-interacted history images, producing fine-grained noise at the feature level. This significantly challenges LMMs to infer implicit user preferences. 

    \item The multimodal instructions may include a reference image alongside textual instructions, \eg ``generate a sticker with the same emotional semantics as this reference image'', requiring LMMs to generate the target image with high-level semantic alignment with the reference image. 
    
    \item Worse still, existing LMMs are not specifically trained for personalized image generation, making it challenging to infer user preferences and align with multimodal instructions. Furthermore, there is a lack of supervised data containing triplets of <user-interacted history images, multimodal instructions, a personalized target image> for LMM training.
    
\end{itemize}

To address the challenges, we propose a \textit{\textbf{P}ersonalized \textbf{I}mage \textbf{Ge}nerati\textbf{on} Framework} (shorted as Pigeon) for LMMs, comprising three key modules:
1) \textit{Mask generation module} incorporates a mask generator to create token-level masks for reference-aware history filtering, effectively removing noisy signals from the history images at the feature level (\cf Section~\ref{sec:mask}). 
2) \textit{Personalization module} integrates masked history tokens and encodes multimodal instructions with the transformed semantic features of the reference image to generate personalized tokens (\cf Section~\ref{sec:personalization}). 
3) \textit{Image generation module} employs a DM to convert the generated personalized tokens into the personalized target image. 

\begin{figure}[t]
\setlength{\abovecaptionskip}{0.1cm}
\setlength{\belowcaptionskip}{-0.45cm}
\centering
\includegraphics[scale=0.55]{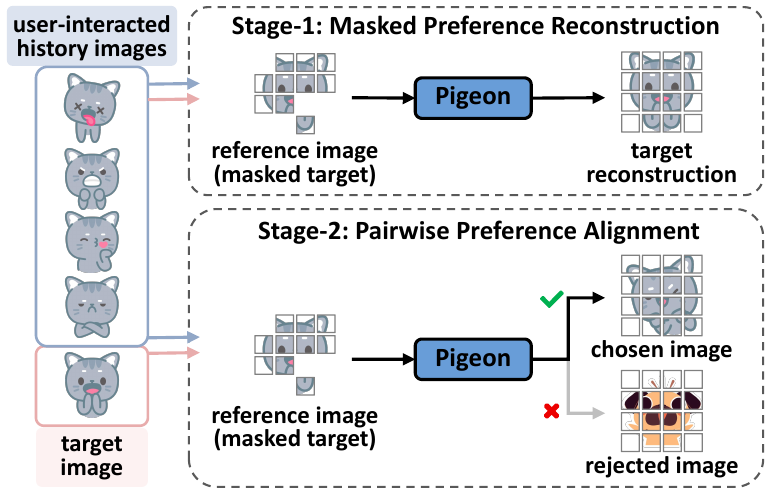}
\caption{Two-stage preference alignments for Pigeon: given user-interacted images, the last image is treated as the target, with the preceding ones as user history.}
\label{fig:alignment}
\end{figure}

Due to the lack of supervised data, Pigeon adopts a two-stage preference alignment scheme to adapt LMMs for personalized image generation.
As shown in Figure~\ref{fig:alignment}, the first stage assumes that user-interacted history images, despite some noise, still partially reflect implicit user preferences. 
Given a sequence of these images, Pigeon treats the last one as the target image and the preceding images as the history images. 
We then mask the target image as a reference image to construct the user's multimodal instructions and fine-tune Pigeon to reconstruct the target image based on this user's history images and multimodal instructions, regulating Pigeon to infer user preference from history. 
In the second stage, Pigeon generates multiple target images based on the first-stage alignment and ranks them using a preference reward strategy, thus forming pseudo-labeled preference data pairs of ``chosen'' and ``rejected'' images. 
Pigeon is then optimized with the preference data pairs via Direct Preference Optimization (DPO)~\cite{rafailov2024direct} to generate more personalized target images, enhancing personalization capabilities. 

\begin{figure*}[t]
\setlength{\abovecaptionskip}{0.0cm}
\setlength{\belowcaptionskip}{-0.4cm}
\centering
\includegraphics[scale=0.48]{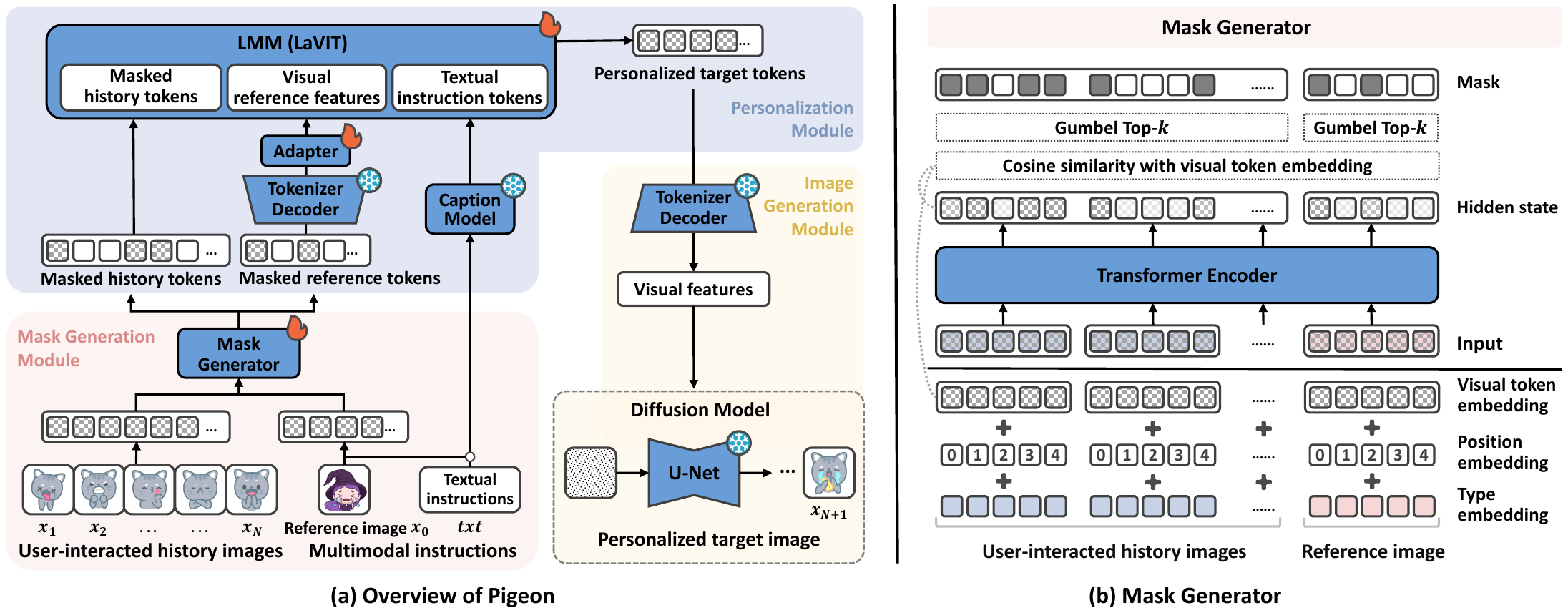}
\caption{Pigeon consists of three key modules: 1) mask generation module creates token-level masks for history and reference images, 2) personalized module encodes multimodal instructions and integrates them with masked history to generate personalized tokens, and 3) image generation module utilizes these tokens to produce personalized images.}
\label{fig:pigeon}
\end{figure*}

We validate the effectiveness of Pigeon in two popular scenarios: personalized sticker and movie poster generation. 
Extensive quantitative evaluation demonstrates that Pigeon outperforms the best baseline in personalization, achieving improvements of 7\%\textasciitilde31\% while maintaining comparable semantic alignment with the reference image. 
Notably, human evaluation on Amazon MTurk\footnote{\url{https://www.mturk.com/}.} reveals that, on average, 71\% participants rate Pigeon-generated images with superior personalization and semantic alignment. 
Furthermore, we discuss the versatility of Pigeon extending to more domains such as personalized product images, advertisement, and fashion images in Appendix~\ref{sec:application}, highlighting Pigeon's broad applicability and significant economic value. 
Our code and data are available at \url{https://github.com/YiyanXu/Pigeon}.

In summary, the key contributions of this work are as follows:
\begin{itemize}[leftmargin=*]
    \item We empower LMMs with the capability of personalized image generation by the Pigeon framework, which can infer user preferences from noisy history images and integrate multimodal instructions for personalized image generation. Pigeon offers a wide range of applications, catering to diverse user demands 
    and driving the evolution of content delivery paradigms.  

    \item We introduce a two-stage preference alignment scheme to effectively adapt LMMs for the personalized image generation task, eliminating the need for supervised data. 

    \item We propose multiple quantitative evaluation metrics for personalized image generation and conduct extensive experiments across two scenarios. Both quantitative results and human evaluation validate that Pigeon significantly surpasses all the baselines, effectively aligning with personalized user preferences. 
    
\end{itemize}


\section{Personalized Image Generation}
In this section, we first formulate the personalized image generation task, followed by the elaboration of our proposed Pigeon framework and its potential applications across various domains.

\subsection{Task Formulation}
Personalized image generation aims to synthesize personalized images tailored to implicit user preferences and explicit multimodal instructions. Formally, given a set of user-interacted history images $\mathcal{H}=\{\bm{x}_i\}_{i=1}^{N}$ and multimodal instructions $\mathcal{R}=\{\bm{x}_0,\bm{txt}\}$, where $\bm{x}_0$ and $\bm{txt}$ represent the reference image and textual instruction, respectively, this goal is to generate a personalized target image $\bm{x}_{N+1}$ that not only meets user visual preferences but also adheres to multimodal instructions by high-level semantic alignment with the reference image. 
This task has broad applications in enhancing user experience across various domains, such as generating personalized product images in e-commerce
or creating personalized movie posters and video thumbnails on platforms like Netflix and YouTube.

\subsection{Pigeon}
\label{sec:method}

To achieve personalized image generation, Pigeon leverages a representative LMM named LaVIT~\cite{lavit} for instantiation\footnote{Pigeon can also be applied to more LMMs, which is left for future exploration.}. Specifically, LaVIT includes a visual tokenizer that translates images into visual tokens for multimodal understanding, and a tokenizer decoder that transforms generated visual tokens into dense visual features to guide image generation. 
Built upon LaVIT, as depicted in Figure \ref{fig:pigeon}(a), Pigeon comprises three key modules: 
1) \textit{mask generation module} employs a mask generator to create token-level masks for both history and reference images.
2) \textit{personalization module} extracts high-level semantic features of multimodal instructions and combines them with the masked history tokens to guide LaVIT to generate personalized tokens that reflect users' content needs.
3) \textit{image generation module} converts these tokens into visual features to generate personalized target images via a DM. 

\subsubsection{\textbf{Mask Generation Module}}
\label{sec:mask}
To discard the noise from user-interacted history images, we introduce a mask generator based on a Transformer encoder~\cite{vaswani2017attention}. It leverages attention mechanisms to encode both history and reference images, and identifies key history tokens that are more relevant to the reference image and contain more personalized information, producing a history mask to filter out noisy tokens.
Besides, the mask generator can also create a token-level mask for the reference image to support the two-stage preference alignments, which will be illustrated in Section~\ref{sec:preference}.

\vspace{3pt}
\noindent\textbf{$\bullet$ Identification of important visual tokens.} 
Given a set of user-interacted history images $\mathcal{H}$ and a reference image $\bm{x}_0$,
we first tokenize these images into visual token sequences:
\begin{equation}
\setlength{\abovedisplayskip}{3pt}
\setlength{\belowdisplayskip}{3pt}
    \bm{E}_i = \text{\textbf{Visual\_Tokenizer}}(\bm{x}_i), i=0,\dots,N,
    \label{eq:tokenize}
\end{equation}
where $\bm{E}_{i}=[\bm{e}_{i1},\dots,\bm{e}_{iL_i}]$ represents the visual token embedding sequence of each image $\bm{x}_i$ with length $L_i$, and $\text{\textbf{Visual\_Tokenizer}}(\cdot)$ refers to the visual tokenizer with a visual embedding layer from the pre-trained LaVIT. 
This process is omitted in Figure~\ref{fig:pigeon}(a) for brevity.
The mask generator, as shown in Figure~\ref{fig:pigeon}(b), combines position and type embeddings with the visual token embeddings via element-wise addition to form the input, which allows the Transformer encoder to distinguish between history and reference tokens and capture the token sequence order within each image. The encoding process is formulated as follows: 
\begin{equation}
\setlength{\abovedisplayskip}{3pt}
\setlength{\belowdisplayskip}{3pt}
    \bm{Z}_1,\dots,\bm{Z}_N,\bm{Z}_0=\text{\textbf{Encoder}}(\bm{E}_1,\dots,\bm{E}_N,\bm{E}_0),
\end{equation}
where $\bm{Z}_i=[\bm{z}_{i1},\dots,\bm{z}_{iL_i}]$ represents the hidden states of each token sequence $\bm{E}_i$, and $\text{\textbf{Encoder}}(\cdot)$ encapsulates both the element-wise addition and the encoding process. 
During the encoding process, the attention mechanism allows the visual tokens from both history and reference images to attend to each other, prioritizing important information while reducing the impact of outlier noise. 
To quantify the importance of each token, we compute the cosine similarity between the hidden states and the original visual token embeddings:
\begin{equation}
\setlength{\abovedisplayskip}{3pt}
\setlength{\belowdisplayskip}{3pt}
    s_{ij} = cosine(\bm{z}_{ij},\bm{e}_{ij}), j=1,\dots,L_i,
    \label{eq:importance}
\end{equation}
where $s_{ij}$ denotes the importance score of the $j$-th token in each visual token sequence $\bm{E}_i$. Intuitively, a higher score indicates more key information is retained in the token. 

\vspace{3pt}
\noindent\textbf{$\bullet$ Reference-aware history filtering.}  
We create a multi-hot binary mask $\bm{m}_h$ to mask the low-score tokens according to the history mask ratio $\alpha_h\in[0,1]$. 
This mask filters out noisy or reference-irrelevant history tokens, yielding the filtered token embeddings for each history image:
$[\bm{\tilde{E}}_1,\dots,\bm{\tilde{E}}_{N}] = \bm{m}_h \odot [\bm{E}_1,\dots,\bm{E}_{N}],$
where $\bm{\tilde{E}}_i$ denotes masked history token embeddings. 
For gradient backpropagation in this discrete sampling process, the Gumbel-Softmax trick~\cite{gumbel} is applied to the non-differentiable binary mask. 

\subsubsection{\textbf{Personalization Module}}
\label{sec:personalization}
To effectively handle multimodal instructions, this module first encodes them to extract essential high-level semantic features, then combines these features with masked history tokens into a hybrid prompt, which serves as the input to LMM, enabling the generation of personalized tokens. 

\vspace{3pt}
\noindent\textbf{$\bullet$ Multimodal instructions encoding.}
When directly utilizing the reference image to guide target image generation, LMMs often duplicate the reference image, failing to effectively incorporate personalized information (see empirical results in Section~\ref{sec:effect_instruction_encoding}). This highlights the necessity to extract high-level semantics from the reference image for image generation.
To enrich the semantics of the reference image $\bm{x}_0$ and enhance the comprehension of multimodal instructions in LMMs, we utilize a caption model (\eg BLIP-2~\cite{li2023blip} and LLaVA~\cite{liu2024llava}) to generate a textual description of the reference image, which is then tokenized into textual tokens: 
\begin{equation}
\setlength{\abovedisplayskip}{3pt}
\setlength{\belowdisplayskip}{3pt}
    \bm{r}_t=\text{\textbf{Text\_Tokenizer}}(\text{\textbf{Caption}}(\bm{x}_0)),
    \label{eq:target_semantics}
\end{equation}
where $\bm{r}_t$ refers to the high-level textual semantic features extracted from the reference image, and $\text{\textbf{Text\_Tokenizer}}(\cdot)$ denotes the text tokenizer with the word embedding layer from LaVIT.

For visual semantics, we transform the low-level reference token embedding sequence $\bm{E}_0$ into high-level dense visual features. Here, we utilize the pre-trained tokenizer decoder of LaVIT for the transform to avoid introducing extra parameters, 
followed by average pooling to aggregate the multiple feature vectors from the tokenizer decoder: 
\begin{equation}
\setlength{\abovedisplayskip}{3pt}
\setlength{\belowdisplayskip}{3pt}
    \bm{v} = \text{\textbf{AvgPooling}}(\text{\textbf{Tokenizer\_Decoder}}(\bm{{E}}_0)).
    \label{eq:visual_feature}
\end{equation}
Next, an adapter layer is introduced to align the feature dimension of $\bm{v}$ with the LaVIT embeddings, \ie $\bm{r}_v = \text{\textbf{Adapter}}(\bm{v})$,
where $\bm{r}_v$ denotes the extracted high-level visual semantic features.

\vspace{3pt}
\noindent\textbf{$\bullet$ Hybrid prompt for LMM.}
To integrate these encoded semantic features with filtered history into prompts for LMM instruction tuning, we propose a hybrid prompt that is structured as follows:
\begin{equation}
\setlength{\abovedisplayskip}{3pt}
\setlength{\belowdisplayskip}{3pt}
    \bm{p} = \text{\textbf{Prompt}}(\bm{\tilde{E}}_1,\dots,\bm{\tilde{E}}_{N}, \bm{r}_t, \bm{r}_v).
    \label{eq:prompt}
\end{equation}

\begin{center}
\begin{tcolorbox}[
                  colbacktitle=gray!30,
                  coltitle=black,
                  colback=white,
                  colframe=black,
                  width=\linewidth,
                  arc=1mm, auto outer arc,
                  boxrule=0.5pt,
                  left=3pt,
                  right=3pt,
                  top=1pt,
                  bottom=-1pt,
                  middle=-1pt,
                  halign title=center,
                 ]
\textbf{Instruction:} You are a helpful personalized assistant. You will receive a list of user-liked images that reflect the user's visual preferences. By analyzing user preferences, please generate a personalized image that aligns with the user's aesthetic taste and the semantics in a specified reference image. 

\textbf{Input:} The user likes the following images: $\mathcolorbox{C7D2E7}{\bm{\tilde{E}}_1,\dots,\bm{\tilde{E}}_{N}}$. The reference image: $\mathcolorbox{F6D7D7}{\bm{r}_t, \bm{r}_v}$.
\tcblower
\textbf{Response:} <Personalized Target Tokens $\mathcolorbox{E7E6E6}{\bm{E}_{N+1}}$>
\end{tcolorbox}
\end{center}
\vspace{-0.1cm}
By using a hybrid prompt similar to the above one, LMMs can adapt to various scenarios to generate personalized target tokens. 

\subsubsection{\textbf{Image Generation Module}}
With personalized target tokens $\bm{E}_{N+1}$, the pre-trained tokenizer decoder of LaVIT converts these discrete tokens into dense visual features, which can guide the generation of the personalized target image $\bm{x}_{N+1}$ in DM. 

\subsubsection{\textbf{Two-stage Preference Alignments}}
\label{sec:preference}
To optimize Pigeon for personalized image generation,
an intuitive strategy is maximizing the generation likelihood of the target tokens $\bm{E}_{N+1}$, based on the prompt $\bm{p}$ in Eq.~(\ref{eq:prompt}). 
However, since there is no supervised dataset containing triplets of <user-interacted history images, multimodal instructions, personalized target image>, we propose a two-stage preference alignment process for effective instruction tuning.

\vspace{3pt}
\noindent\textbf{$\bullet$ Stage-1: Masked Preference Reconstruction.} 
We assume that user-interacted history images, despite containing some noise, still reflect user visual preferences. Based on this, as shown in Figure~\ref{fig:alignment}, given a sequence of user-interacted images $\{\bm{x}_i\}_{i=1}^{N+1}$, the last one $\bm{x}_{N+1}$ is considered the personalized target image, while the preceding images are treated as history images $\mathcal{H}=\{\bm{x}_i\}_{i=1}^{N}$. 

\vspace{3pt}
\textbf{Supervised dataset construction.}
Considering the lack of multimodal instructions, we adopt the target image as the reference to construct multimodal instructions $\mathcal{R}=\{\bm{x}_{N+1},\bm{txt}\}$. A token-level reference mask is then applied to corrupt the reference image, encouraging the model to extract user preferences from history images for target reconstruction. Specifically, we utilize the importance score defined in Eq.~(\ref{eq:importance}) to rank all the reference tokens and create the token-level mask for the reference image.

Unlike the history mask, which filters out noise by discarding low-score tokens, we introduce a dual-phase mask scheme for the reference image. During training, we mask high-score reference tokens, which contain more personalized information (as discussed in Section~\ref{sec:mask}), forcing the model to rely on history images to recover the target. 
During inference, low-score tokens are masked instead, utilizing the preference reconstruction capability to generate more personalized content. 
Formally, the dual-phase mask $\bm{m}_r$ with a reference mask ratio $\alpha_r\in[0,1]$ is applied to the reference tokens 
by $\tilde{ \bm{E}}_0=\bm{m}_r\odot\bm{E}_0$. 
We then replace $\bm{E}_0$ in Eq.~(\ref{eq:visual_feature}) with
$\tilde{ \bm{E}}_0$ to derive the modified visual features for the hybrid prompt $\bm{p}$ in Eq.~(\ref{eq:prompt}), optimizing the model to reconstruct the target token sequence $\bm{E}_{N+1}$. 
In this way, we could construct a supervised prompt-response dataset $\mathcal{D}=\{(\bm{p}, \bm{E}_{N+1})^k\}_k$ from the available interaction sequences for masked preference reconstruction. 

\vspace{3pt}
\textbf{Supervised fine-tuning.}
For parameter-efficient fine-tuning, we introduce a LoRA~\cite{hu2022lora} module into the pre-trained LaVIT, which keeps the LaVIT parameters frozen and imports trainable low-rank decomposition matrices for updates. As shown in Figure~\ref{fig:pigeon}(a), we only fine-tune specific components of Pigeon, namely the mask generator, adapter, and LoRA for LaVIT, while freezing all the other parameters. During training, we randomly sample the reference mask ratio $\alpha_r\in[0,1]$ and fine-tune Pigeon for target reconstruction, aiming to capture more robust user preferences. Formally, the loss function is defined as the negative likelihood of the target token sequence via an auto-regressive manner:
\begin{equation}
    \resizebox{.9\hsize}{!}{$\mathcal{L}_{sft} = -\sum_{\substack{(\bm{p},\bm{E}_{N+1})\in\mathcal{D}\\\alpha_r\sim\mathcal{U}(0,1)}}\sum_{j=1}^{L_{N+1}}\log\left( P_{\Theta}(\bm{e}_{N+1,j}|\bm{p}(\alpha_r),\bm{e}_{N+1,<j})\right)$}, 
\end{equation}
where $\bm{e}_{N+1,j}$ is the $j$-th token in the sequence $\bm{E}_{N+1}$ of length $L_{N+1}$, $\bm{p}(\alpha_r)$ is the hybrid prompt with a uniformly sampled reference mask ratio, and $\bm{\Theta}$ includes all the learnable parameters of Pigeon.

\vspace{3pt}
\noindent\textbf{$\bullet$ Stage-2: Pairwise Preference Alignment.} 
After the first-stage fine-tuning,
Pigeon is capable of following the instructions for personalized image generation. To further enhance its personalization capability, we adopt DPO~\cite{rafailov2024direct} for pairwise preference alignment, which utilizes preference pairs of chosen and rejected responses to optimize the model to produce the chosen one. 

\vspace{3pt}
\textbf{Preference dataset construction.}
To construct the preference data pairs for DPO, we first generate multiple personalized target token sequences for each prompt $\bm{p}(\alpha_r)$ with varying reference mask ratios $\alpha_r\in\{0.0,0.1,\dots,1.0\}$ based on the first-stage alignment. These tokens are then transformed into images $\bm{x}(\alpha_r)$ via the image generation module. To identify the best and worst personalized images, we introduce a preference reward strategy to rank all generated images. Following~\cite{pmg}, we compute the CLIP similarity between each generated image and the history images:
\begin{equation}
\setlength{\abovedisplayskip}{3pt}
\setlength{\belowdisplayskip}{3pt}
    s(\alpha_r) = \frac{1}{N}{\textstyle\sum_{i=1}^{N}}\text{\textbf{CLIPSim}}(\bm{x}(\alpha_r),\bm{x}_i),
\label{eq:preference_score}
\end{equation}
where $s(\alpha_r)$ is the preference score of image $\bm{x}(\alpha_r)$. 
We rank the generated images based on these scores to form the pseudo-labeled preference dataset $\bar{\mathcal{D}}=\{(\bm{p}, \bm{E}', \bm{E}'')^k\}_k$, where $\bm{E}'$ and $\bm{E}''$ denote the chosen and rejected token sequences for DPO, corresponding to images with the highest and lowest preference scores.

\vspace{3pt}
\textbf{Preference optimization.}
In this stage, we continue updating the LoRA weights while keeping all the other parameters frozen. With the preference dataset, the loss function can be formulated as:
\begin{equation}
    \resizebox{.9\hsize}{!}{$\mathcal{L}_{DPO}=-\mathbb{E}_{\substack{(\bm{p},\bm{E}',\bm{E}'')\sim\mathcal{\bar{D}}\\\alpha_r\sim\mathcal{U}(0,1)}}\left[\log\sigma\left(\beta\dfrac{P_{\Theta_l}(\bm{E}'|\bm{p}(\alpha_r))}{{P_{\hat{\Theta}_l}}(\bm{E}'|\bm{p}(\alpha_r))}-\beta\dfrac{P_{\Theta_l}(\bm{E}''|\bm{p}(\alpha_r))}{{P_{\hat{\Theta}_l}}(\bm{E}''|\bm{p}(\alpha_r))}\right)\right]$},
\end{equation}
where ${\Theta}_l$ denotes the learnable parameters of the LoRA module, and $\beta$ is a parameter controlling the deviation from the reference model $\hat{\Theta}_l$ obtained in the first-stage alignment. 

\subsubsection{\textbf{Inference}}
To manage the trade-off between personalization and semantic alignment with the reference image, users could adjust the reference mask ratio to control how much reference information is retained in the generated images. 
During inference, given history images $\mathcal{H}=\{\bm{x}_i\}_{i=1}^N$, multimodal instructions $\mathcal{R}=\{\bm{x}_0,\bm{txt}\}$ and a user-specified reference mask ratio $\alpha_r$, 
Pigeon can mask the low-score reference tokens accordingly to generate an image $\bm{x}_{N+1}$ that aligns with the user's visual preferences and multimodal instructions.

\section{Experiments}
\label{sec:experiment}

\begin{table*}[t]
\setlength{\abovecaptionskip}{0.05cm}
\setlength{\belowcaptionskip}{0.2cm}
\caption{Quantitative performance comparison between Pigeon and the baselines in both scenarios. Baselines labeled with ``$\ast$" indicate the pre-trained models. The best results are highlighted in bold, while the second-best results are underlined.}
\label{table:overall}
\resizebox{0.98\textwidth}{!}{
\begin{tabularx}{\textwidth}{
XX|
>{\centering\arraybackslash\hsize=1.2\hsize\linewidth=\hsize}X|
>{\centering\arraybackslash\hsize=.9\hsize\linewidth=\hsize}X
>{\centering\arraybackslash\hsize=.9\hsize\linewidth=\hsize}X
>{\centering\arraybackslash\hsize=.9\hsize\linewidth=\hsize}X
>{\centering\arraybackslash}X
>{\centering\arraybackslash\hsize=1.4\hsize\linewidth=\hsize}X|
>{\centering\arraybackslash\hsize=.9\hsize\linewidth=\hsize}X
>{\centering\arraybackslash\hsize=.9\hsize\linewidth=\hsize}X
>{\centering\arraybackslash\hsize=.9\hsize\linewidth=\hsize}X|
>{\centering\arraybackslash}X}
\hline
\multicolumn{2}{l|}{\textbf{\#Sticker}}                                                             & \multicolumn{1}{l|}{}                  & \multicolumn{5}{c|}{\textbf{Personalization}}                                                                                                                                                                & \multicolumn{3}{c|}{\textbf{Semantic Alignment}}                                              & \textbf{Fidelity}             \\
\multicolumn{2}{c|}{\textbf{Methods}}                                                               & \textbf{Overall}              & \textbf{CS$\uparrow$}                           & \textbf{CIS$\uparrow$}                          & \textbf{DIS$\uparrow$}                          & \textbf{LPIPS$\downarrow$}                         & \textbf{MS-SSIM$\uparrow$}                       & \textbf{CS$\uparrow$}                  & \textbf{CIS$\uparrow$}                 & \textbf{DIS$\uparrow$}                 & \textbf{FID$\downarrow$}                 \\ \hline
\multicolumn{1}{l}{\textbf{DM-based}}                    & \textbf{TI}              & {\myul 36.91}                            & 18.67                                  & 40.90                                  & 36.58                                  & 0.7654                                  & 0.0887                                  & \textbf{32.91}                & {\myul 53.67}                   & {\myul 48.50}                   & 105.48                        \\ \hline
\multicolumn{1}{l}{\textbf{LLM-based}}                   & \textbf{PMG}                            & 32.83                                  & {\myul 19.16}                            & 47.34                                  & 39.15                                  & 0.7383                                  & 0.0827                                  & 18.31                         & 45.45                         & 37.80                         & {\myul 84.91 }                        \\ \hline
\multicolumn{1}{c}{}                                     & \textbf{LLaVA*}                         & 32.40                                  & 17.88                                  & 47.26                                  & 42.59                                  & 0.7575                                  & 0.0966                                  & 17.54                         & 42.65                         & 39.25                         & 93.23                         \\
\multicolumn{1}{c}{}                                     & \textbf{LLaVA}                          & 32.23                                  & 18.72                                  & 37.44                                  & 33.19                                  & 0.7552                                  & 0.0851                                  & {\myul 27.02}                   & 49.15                         & 43.88                         & 95.19                         \\
\multicolumn{1}{c}{}                                     & \textbf{LaVIT*}                         & 34.56                                  & 18.77                                  & {\myul 53.63}                            & {\myul 50.96}                            & {\myul 0.6855}                            & {\myul 0.1376}                            & 15.49                         & 40.76                         & 39.09                         & 107.53                        \\
\multicolumn{1}{c}{}                                     & \textbf{LaVIT}                          & 33.15                                  & 16.39                                  & 40.56                                  & 40.84                                  & 0.7377                                  & 0.1128                                  & 25.74                         & \textbf{70.80}                & \textbf{69.93}                & \textbf{83.39}                \\
\multicolumn{1}{l}{\multirow{-5}{*}{\textbf{LMM-based}}} & \cellcolor[HTML]{EAEAEA}\textbf{Pigeon} & \cellcolor[HTML]{EAEAEA}\textbf{44.38} & \cellcolor[HTML]{EAEAEA}\textbf{23.69} & \cellcolor[HTML]{EAEAEA}\textbf{67.65} & \cellcolor[HTML]{EAEAEA}\textbf{62.23} & \cellcolor[HTML]{EAEAEA}\textbf{0.6814} & \cellcolor[HTML]{EAEAEA}\textbf{0.1568} & \cellcolor[HTML]{EAEAEA}21.10 & \cellcolor[HTML]{EAEAEA}47.44 & \cellcolor[HTML]{EAEAEA}45.44 & \cellcolor[HTML]{EAEAEA}89.43 \\ \hline
\multicolumn{1}{c}{}                                     & \textbf{Recon}                    & 33.22                                  & 16.30                                  & 40.60                                  & 40.76                                  & 0.7370                                  & 0.1126                                  & 25.84                         & 71.09                         & 70.14                         & 83.57                         \\
\multicolumn{1}{l}{\multirow{-2}{*}{\textbf{Reference}}} & \textbf{Grd}                            & 36.98                                  & 16.93                                  & 45.00                                  & 43.71                                  & 0.6443                                  & 0.1349                                  & 28.95                         & 100.00                        & 100.00                        & -                             \\ \hline
\end{tabularx}
}

\resizebox{0.98\textwidth}{!}{
\begin{tabularx}{\textwidth}{
XX|
>{\centering\arraybackslash\hsize=1.2\hsize\linewidth=\hsize}X|
>{\centering\arraybackslash\hsize=.9\hsize\linewidth=\hsize}X
>{\centering\arraybackslash\hsize=.9\hsize\linewidth=\hsize}X
>{\centering\arraybackslash\hsize=.9\hsize\linewidth=\hsize}X
>{\centering\arraybackslash}X
>{\centering\arraybackslash\hsize=1.4\hsize\linewidth=\hsize}X|
>{\centering\arraybackslash\hsize=.9\hsize\linewidth=\hsize}X
>{\centering\arraybackslash\hsize=.9\hsize\linewidth=\hsize}X
>{\centering\arraybackslash\hsize=.9\hsize\linewidth=\hsize}X|
>{\centering\arraybackslash}X}
\hline
\multicolumn{2}{l|}{\textbf{\#Movie poster}}                                                        & \multicolumn{1}{l|}{}                  & \multicolumn{5}{c|}{\textbf{Personalization}}                                                                                                                                                             & \multicolumn{3}{c|}{\textbf{Semantic Alignment}}                                                          & \textbf{Fidelity}             \\
\multicolumn{2}{c|}{\textbf{Methods}}                                                               & \textbf{Overall}                       & \textbf{CS$\uparrow$}                           & \textbf{CIS$\uparrow$}                          & \textbf{DIS$\uparrow$}                       & \textbf{LPIPS$\downarrow$}                         & \textbf{MS-SSIM$\uparrow$}                       & \textbf{CS$\uparrow$}                  & \textbf{CIS$\uparrow$}                       & \textbf{DIS$\uparrow$}                       & \textbf{FID$\downarrow$}                 \\ \hline
\multicolumn{1}{l}{\textbf{DM-based}}                    & \textbf{TI}                             & {\myul 31.07}                            & 12.41                                  & 28.29                                  & 19.18                               & 0.7721                                  & 0.0399                                  & {\textbf{33.84}}                   & 43.53                               & 39.81                               & 79.77                         \\ \hline
\multicolumn{1}{l}{\textbf{LLM-based}}                   & \textbf{PMG}                            & 20.36                                  & 13.61                                  & 25.11                                  & \textbf{22.73}                      & 0.7692                                  & 0.0261                                  & 15.60                         & 27.29                               & 25.15                               & 77.25                         \\ \hline
\multicolumn{1}{c}{}                                     & \textbf{LLaVA*}                         & 22.08                                  & 12.24                                  & 29.60                                  & 19.73                               & 0.7607                                  & 0.0373                                  & 14.55                         & 31.76                               & 21.99                               & 73.77                         \\
\multicolumn{1}{c}{}                                     & \textbf{LLaVA}                          & 30.59                                  & 12.62                                  & {\myul 30.64}                                  & 19.33                               & 0.7690                                  & 0.0370                                  & {\myul 30.53}                         & {\myul 48.50}                               & 41.45                               & 54.55                         \\
\multicolumn{1}{c}{}                                     & \textbf{LaVIT*}                         & 23.81                                  & 12.64                                  & 28.23                                  & 17.50                               & {\myul 0.7546}                            & {\myul 0.0458}                            & 19.39                         & 36.93                               & 37.71                               & 50.08                         \\
\multicolumn{1}{c}{}                                     & \textbf{LaVIT}                          & 27.82                                  & {\myul 13.86}                            & 30.49                                  & 19.95                               & 0.7548                                  & 0.0370                                  & 25.15                         & 46.02                               & \textbf{60.07}                      & \textbf{33.53}                \\
\multicolumn{1}{l}{\multirow{-5}{*}{\textbf{LMM-based}}} & \cellcolor[HTML]{EAEAEA}\textbf{Pigeon} & \cellcolor[HTML]{EAEAEA}\textbf{33.31} & \cellcolor[HTML]{EAEAEA}\textbf{15.41} & \cellcolor[HTML]{EAEAEA}\textbf{40.16} & \cellcolor[HTML]{EAEAEA}{\myul 21.29} & \cellcolor[HTML]{EAEAEA}\textbf{0.7508} & \cellcolor[HTML]{EAEAEA}\textbf{0.0464} & \cellcolor[HTML]{EAEAEA}26.45 & \cellcolor[HTML]{EAEAEA}{\textbf{49.66}} & \cellcolor[HTML]{EAEAEA}{\myul 44.07} & \cellcolor[HTML]{EAEAEA}{\myul
47.79} \\ \hline
\multicolumn{1}{c}{}                                     & \textbf{Recon}                          & 27.81                                  & 13.85                                  & 30.33                                  & 19.95                               & 0.7548                                  & 0.0367                                  & 25.29                         & 46.08                               & 60.52                               & 33.74                         \\
\multicolumn{1}{l}{\multirow{-2}{*}{\textbf{Reference}}} & \textbf{Grd}                            & 41.58                                  & 10.94                                  & 51.34                                  & 20.75                               & 0.7502                                  & 0.0402                                  & 31.81                         & 100.00                              & 100.00                              & -                             \\ \hline
\end{tabularx}
}
\vspace{-0.35cm}
\end{table*}

We evaluate Pigeon in sticker and movie poster scenarios to validate its superiority by answering the following research questions:
\begin{itemize}[leftmargin=*]
    \item \textbf{RQ1:} How does Pigeon perform compared with DM-based, LLM-based, and LMM-based personalized image generation methods, based on quantitative evaluation?

    \item \textbf{RQ2:} Can Pigeon surpass the baselines in human evaluation?

    \item \textbf{RQ3:} How do the special designs of Pigeon (\eg history mask, multimodal instruction encoding strategy, and two-stage preference alignment process) affect the performance?
\end{itemize}

\subsection{Experimental Settings}
\subsubsection{\textbf{Datasets}} 
We conduct experiments on two datasets, focusing on sticker and movie poster scenarios: 1) \textbf{SER30K}\footnote{https://github.com/nku-shengzheliu/SER30K.} is a large-scale dataset of stickers, each categorized by theme and annotated with an associated emotion label; and 2) \textbf{ML-Latest}\footnote{https://grouplens.org/datasets/movielens.}, a benchmark dataset containing user ratings on movies. Details regarding data processing and dataset statistics are available in Appendix~\ref{sec:appendix_data}.


\subsubsection{\textbf{Baselines}} 
We compare Pigeon with various generative baselines, including methods based on DMs, LLMs, and LMMs: \textbf{1) Textual Inversion (TI)}~\cite{textualinversion} introduces a word embedding to learn user preference representation, which is then combined with textual instructions to guide the text-to-image generation process in DMs. \textbf{2) PMG}~\cite{pmg} transforms user-interacted and reference images into textual descriptions, using pre-trained LLMs to extract user preferences through keywords and implicit embeddings to condition the image generator. \textbf{3) LLaVA}~\cite{liu2024llava} is an LMM designed to extract dense image features for visual reasoning, generating text by default but capable of producing images when integrated with an external text-to-image generator. \textbf{4) LaVIT}~\cite{lavit} is another LMM that converts images into discrete visual tokens for reasoning and generates visual tokens to guide the image generation process.

Additionally, we include two results for reference: 5) \textbf{Recon}, which utilizes the visual tokenizer, tokenizer decoder, and DM of the pre-trained LaVIT for image reconstruction without personalization; and 6) \textbf{Grd}, representing the evaluation results of the reference images. The performance gap between Recon and Grd reflects the difference between generated and real-world images. 

\subsubsection{\textbf{Evaluation Metrics}}
We employ various quantitative evaluation metrics for performance comparison. Following~\cite{pmg,shilova2023adbooster}, we mainly focus on \textbf{personalization} and \textbf{semantic alignment} with the reference image by measuring the semantic and perceptual similarity between generated and history/reference images. 
\begin{itemize}[leftmargin=*]
    \item \textbf{Semantic similarity.} We adopt CLIP~\cite{radford2021learning} and DINO~\cite{oquab2024dinov}
    to extract image features from generated and history/reference images, and compute the cosine similarity between them to obtain the CLIP Image Score (CIS) and DINO Image Score (DIS). Additionally, the CLIP Score (CS) measures the similarity between generated images and textual descriptions of the history/reference images. To assess the overall performance, we also calculate a unified F1-score, combining the history CIS and reference CS.

    \item \textbf{Perceptual similarity.} To evaluate finer-grained visual personalization, we apply LPIPS~\cite{zhang2018unreasonable} and MS-SSIM~\cite{wang2003multiscale} to quantify the perceptual similarity between generated and history images.

    \item \textbf{Fidelity.} 
    We also employ the widely-used FID metric to assess the fidelity of the generated images.
\end{itemize}

\subsubsection{\textbf{Implementation Details}} All the baselines are tuned with a fixed learning rate of $1e^{-5}$. We implement PMG following its default model designs, while other baselines are implemented with Stable Diffusion XL~\cite{podell2024sdxl} as the image generator for fair comparisons. Detailed hyper-parameter
settings and computational overhead of Pigeon are summarized in Appendix~\ref{sec:appendix_cost}.



\subsection{Quantitative Evaluation (RQ1)}
The comparison between Pigeon and the baselines is shown in Table~\ref{table:overall}. The observations are summarized as follows:
\begin{itemize}[leftmargin=*]
    \item DM-based TI outperforms most baselines in semantic alignment by directly using the textual description of the reference image for text-to-image generation. However, noisy signals in interaction history hinder its ability to precisely capture user preferences, resulting in inferior personalization.

    \item PMG converts images into textual descriptions and uses LLMs to infer user preferences for guiding image generation. The image-to-text conversion may overlook critical visual details, leading to inaccurate preference modeling and multimodal instruction understanding. As a result, PMG presents moderate performances in both personalization and semantic alignment.

    \item The decent performance of the pre-trained LLaVA and LaVIT in personalization validates the strength of advanced instruction-following and visual understanding capabilities in LMMs for personalized image generation. 
    Among them, LLaVA relies on personalized text to guide image generation, which can cause misalignments between expressed textual preferences and actual visual preferences, resulting in relatively lower performance.

    \item After fine-tuning in each scenario, both LLaVA and LaVIT tend to reconstruct reference images rather than generate personalized ones, as evidenced by significant improvements in semantic alignment alongside a decline in personalization. This is mainly due to the lack of supervised data for model training.

    \item Pigeon exhibits superior performance in most personalization metrics across two scenarios, while maintaining comparable semantic alignment and fidelity. These results underscore the effectiveness of Pigeon in capturing user visual preferences from noisy history images and accurately understanding multimodal instructions to produce personalized images.
\end{itemize}

\begin{figure*}[t]
\setlength{\abovecaptionskip}{0.cm}
\setlength{\belowcaptionskip}{-0.4cm}
\centering
\includegraphics[scale=0.467]{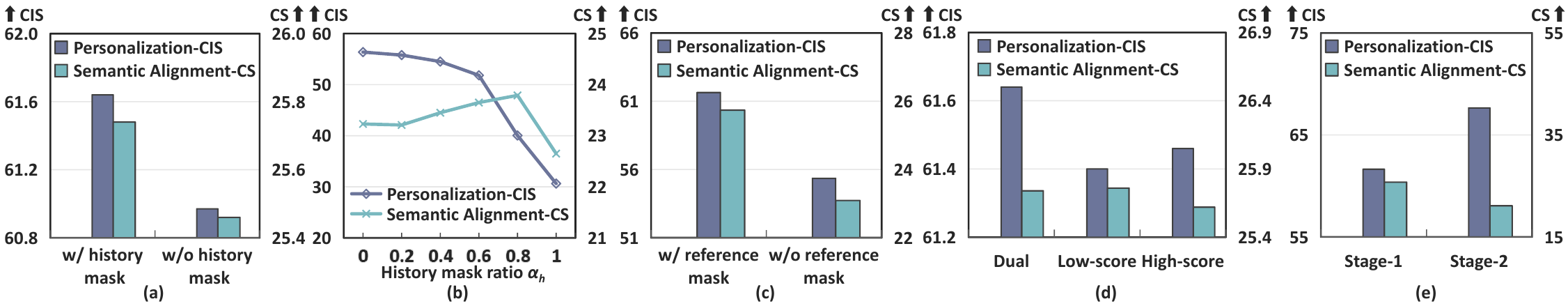}
\caption{In-depth analysis of the history mask and the two-stage preference alignment process.}
\label{fig:mask}
\end{figure*}

\subsection{Human Evaluation (RQ2)}
To assess the qualitative performance of Pigeon in personalization and semantic alignment, we conduct a human evaluation on Amazon MTurk\footnote{https://www.mturk.com/.}, comparing it against Grd and two representative baselines: 1) TI, which exhibits the second-best overall performance in Table~\ref{table:overall} , and 2) PMG, designed for personalized image generation. The evaluation adopts binary-choice tests across sticker and movie poster scenarios, each with 50 cases. For personalization, we present five user-interacted history images and the generated images, with the question: ``When provided with someone's five previously liked stickers (movies), please select the next sticker (movie poster) that is more attractive to her/him.'' For semantic alignment, we display the reference and generated images with the question: ``Which image aligns more closely with the semantics of the reference image?'' 
As shown in Table~\ref{table:human_eval},
Pigeon consistently surpasses ($\ge50\%$) the baselines, even the Grd, in personalization and maintains decent results in semantic alignment with reference images. These findings emphasize its superiority in capturing user preferences from noisy history images and effectively integrating multimodal instructions for image generation,
which aligns with the quantitative analysis. More detailed information can be found in Appendix~\ref{sec:appendix_human_eval}.

\begin{table}[t]
\setlength{\abovecaptionskip}{0.05cm}
\setlength{\belowcaptionskip}{0cm}
\caption{The human evaluation results, where ``$\pm$'' denotes 95\% confidence interval. Pigeon is consistently preferred ($\ge50\%$) over the baselines across sticker and movie poster scenarios.}
\label{table:human_eval}
\resizebox{0.48\textwidth}{!}{
\begin{tabular}{cl|ccc}
\hline
\multicolumn{2}{c|}{\textbf{Pigeon}}                                                                      & \textbf{Grd} & \textbf{TI} & \textbf{PMG} \\ \hline\xrowht{10pt}
\multirow{2}{*}{\textbf{Personalization}}                                              & \textbf{Sticker} & 0.91$^{\pm2.19\%}$  & 0.91$^{\pm2.19\%}$ & 0.89$^{\pm1.79\%}$  \\
& \textbf{Movie}   & 0.62$^{\pm2.85\%}$  & 0.66$^{\pm2.16\%}$ & 0.57$^{\pm2.51\%}$  \\ \hline\xrowht{10pt}
\multirow{2}{*}{\textbf{\begin{tabular}[c]{@{}c@{}}Semantic\\ Alignment\end{tabular}}} & \textbf{Sticker} & -            & 0.54$^{\pm2.67\%}$ & 0.67$^{\pm3.65\%}$  \\
& \textbf{Movie}   & -            & 0.58$^{\pm2.58\%}$ & 0.73$^{\pm2.22\%}$  \\ \hline
\end{tabular}
}
\vspace{-0.4cm}
\end{table}

\subsection{In-depth Analysis (RQ3)}
In this section, we conduct additional experiments in the sticker scenario to further investigate the effects of various Pigeon designs, including the history mask, multimodal instruction encoding strategy, and the two-stage preference alignment process. To reduce resource costs, we mainly focus on the results after first-stage preference alignment for fair comparisons.

\subsubsection{\textbf{Effect of history mask}}
To assess the effectiveness of the history mask in managing noisy history images, we exclude it during training and present the results on two key metrics in Figure~\ref{fig:mask}(a).
The findings show that:
1) noise in the history images prevents the model from accurately capturing user preferences and even disrupts the semantic alignment with the reference image.
2) The history mask could effectively filter out the noisy signals, thereby enhancing model performance.

Additionally, we vary the history mask ratio $\alpha_h$ during inference, with the reference mask ratio fixed at 0.5. The results in Figure~\ref{fig:mask}(b) reveal that increasing $\alpha_h$ discards both noise and useful personalized information in history images, causing Pigeon to rely more on the reference image, thus slightly improving the semantic alignment. However, this also makes it harder for Pigeon to extract user preferences, reducing the performance in personalization.

\begin{table}[t]
\setlength{\abovecaptionskip}{0.05cm}
\setlength{\belowcaptionskip}{0cm}
\caption{Effects of multimodal instruction encoding.}
\label{table:encoding}
\resizebox{0.48\textwidth}{!}{
\begin{tabular}{l|cc|c}
\hline
                            & \multicolumn{2}{c|}{\textbf{Personalization}} & \textbf{Semantic Alignment} \\
                            & \textbf{CIS$\uparrow$}          & \textbf{LPIPS$\downarrow$}        & \textbf{CS$\uparrow$}                 \\ \hline
\textbf{Pigeon}             & \textbf{61.64}        & \textbf{0.6800}       & \textbf{25.74}              \\
\textbf{- w/o visual feature} & 55.46                 & 0.6828                & 23.53                       \\
\textbf{- w/o textual tokens} & 65.73                 & 0.6731                & 20.37                       \\
\textbf{- w/o encoding}       & 55.35                 & 0.6976                & 24.72                       \\ \hline
\end{tabular}
}
\vspace{-0.3cm}
\end{table}

\subsubsection{\textbf{Effect of multimodal instruction encoding}}
\label{sec:effect_instruction_encoding}
To validate the necessity to extract high-level semantics via multimodal instruction encoding, we perform three ablation studies during the training phase. Specifically, we remove the encoded visual features and textual instruction tokens, referred to as ``w/o visual features'' and ``w/o textual tokens'', respectively. We also disable the encoding process by directly inputting the masked reference tokens into LaVIT, denoted as ``w/o encoding''. Results on three key metrics, reported in Table~\ref{table:encoding}, reveal the following insights: 
1) removing the visual features reduces the performance, highlighting the importance of high-level visual semantics for understanding the reference image and enhancing personalization. 
2) Excluding textual tokens improves personalization while significantly reducing semantic alignment, indicating that the model over-prioritizes user preferences when textual semantics are absent.
3) Disabling the encoding process leads to simple duplication of the reference image rather than true personalization, as evidenced by a notable drop in personalization and an increase in semantic alignment.

\subsubsection{\textbf{Effect of two-stage preference alignments}}
\leavevmode

\vspace{3pt}
\noindent\textbf{$\bullet$ Stage-1: masked preference reconstruction.} 
To evaluate the impact of the first-stage masked preference reconstruction, we perform additional experiments that analyze the effect of the reference mask and explore alternative masking schemes: 1) removing the reference mask, as shown in Figure~\ref{fig:mask}(c), leads to a notable performance decline, underscoring the importance of masked preference reconstruction, which allows Pigeon to effectively integrate user preferences with reference semantics for personalization. 2) Exploring alternative masking schemes for the reference tokens: ``Low-score'' refers to masking low-score tokens during both training and inference, while ``High-score'' masks high-score tokens in both phases. These schemes are compared to the dual mask scheme of Pigeon, with results presented in Figure~\ref{fig:mask}(d). The significant decline in personalization suggests that masking either high-score or low-score tokens during both phases causes the model to over-focus on preference reconstruction, limiting its ability to generalize this reconstruction for broader personalization.

\vspace{3pt}
\noindent\textbf{$\bullet$ Stage-2: pairwise preference alignment.}
We evaluate the effect of the second-stage pairwise preference alignment by comparing the performance after the first and second stages of alignments, as shown in Figure~\ref{fig:mask}(e). Despite a slight decline in semantic alignment, the second-stage preference alignment further enhances personalization. This demonstrates the effectiveness of DPO in aligning the generation process more closely with user preferences, ultimately resulting in more personalized image generation. Analysis of different preference reward strategies during the second-stage alignment is presented in Appendix~\ref{sec:appendix_reward}.

\begin{figure}[t]
\setlength{\abovecaptionskip}{0.2cm}
\setlength{\belowcaptionskip}{-0.3cm}
\centering
\includegraphics[scale=0.485]{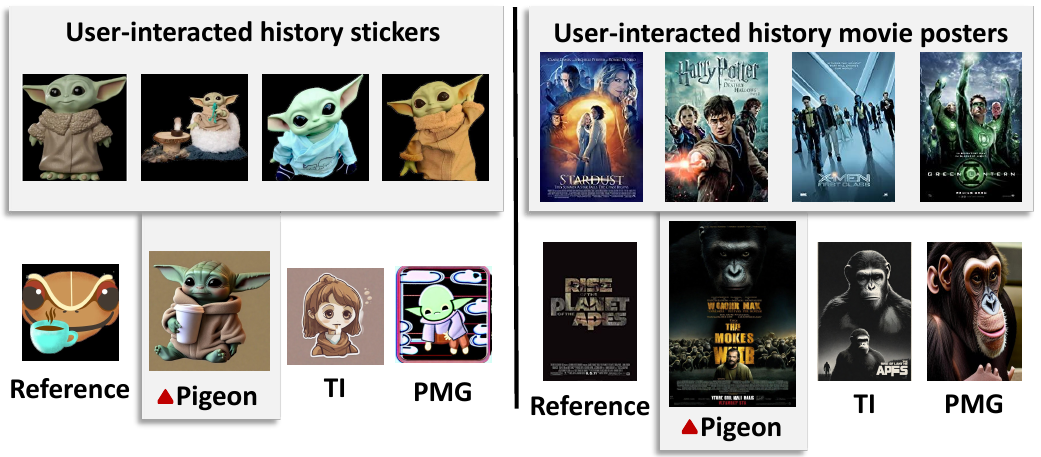}
\caption{Examples of generated images in sticker and movie poster scenarios, along with four user-interacted history images and one reference image.}
\label{fig:case}
\end{figure}


\subsection{Case Study}
\label{sec:case_study}
In this section, we present two examples of Pigeon-generated images in sticker and movie poster scenarios, along with four user-interacted history images and one reference image. We compare Pigeon with two competitive baselines, TI and PMG, as shown in Figure~\ref{fig:case}. In the sticker scenario, Pigeon effectively captures the user's visual preference for Yoda and integrates it with the high-level semantics of the reference sticker, such as ``drinking coffee'', achieving impressive personalization and semantic alignment with the reference image. In the movie poster scenario, Pigeon-generated poster for the movie ``Rise of the Planet of the Apes'' showcases high semantic alignment with the reference poster by emphasizing an intense central ape figure, evoking a similar sense of power and conflict. Meanwhile, it matches the user's preference for character-centered movie posters with a dark and dramatic color palette. More examples are provided in Figure~\ref{fig:sticker_cases} and Figure~\ref{fig:movie_cases} in Appendix~\ref{sec:appendix_case}.


\section{Related Work}
\label{sec:related_work}

\noindent\textbf{$\bullet$ Personalized Content Filtering.}
Traditional filtering-based personalized content delivery approaches, such as recommender systems~\cite{wang2023diffusion,lin2024bridging,bao2023tallrec,zhao2024denoising,ClusterGCF2024TOIS}, typically rank existing content based on user interaction history and contextual information, delivering the top-ranked content. However, constrained by the limited diversity of available content, they often fall short of meeting users' diverse needs~\cite{xu2024diffusion,yu2019personalized,wang2023generative}, motivating the emergence of personalized content generation across various domains.

\vspace{3pt}
\noindent\textbf{$\bullet$ Personalized Content Generation.}
The rise of powerful generative models, such as DMs~\cite{podell2024sdxl,rombach2022high}, LLMs~\cite{touvron2023llama}, and LMMs~\cite{liu2024llava,lavit}, has sparked increasing interest in their potential for personalized content generation. Most previous work focuses on personalized text generation~\cite{kumar2024longlamp,salemi2024optimization,rame2024rewarded,tan2024personalized}. 
For example, the LaMP benchmark~\cite{salemi-etal-2024-lamp} is developed to train and evaluate LLMs in various personalized text scenarios like personalized news headline generation and tweet paraphrasing. Further work, such as RSPG~\cite{salemi2024optimization}, studies the retrieval-augmented solutions to personalize LLM outputs, while PER-PCS~\cite{tan2024personalized} introduces a parameter-sharing framework to enable more efficient and fine-grained personalization.

In contrast, personalized image generation has received relatively less attention. Current research mainly adopts DMs and LLMs for this task: 
1) DM-based methods~\cite{li2024stylegan,customdiff,he2024imagine,shi2024instantbooth}, such as TI~\cite{textualinversion} and DreamBooth~\cite{ruiz2023dreambooth}, focus on aligning image generation with users' explicit multimodal instructions, without consideration of user implicit visual preferences.  Other approaches like DiFashion~\cite{xu2024diffusion}, CG4CTR~\cite{yang2024new}, and AdBooster~\cite{shilova2023adbooster}, integrate user data (\eg interaction history and user features) with multimodal instructions to guide personalized fashion and product image generation. However, these methods often struggle with the noisy signals in user-interacted history images, leading to inaccurate preference modeling. 
2) LLM-based PMG~\cite{pmg} translates images into texts for the LLM to extract user visual preferences, while the limitations of text in conveying complex visual details hinder its effectiveness. 
In this work, we leverage the notable visual understanding and reasoning capabilities of LMMs, along with dedicated modules, to develop the Pigeon framework that effectively handles noisy history images for accurate, tailored image generation. 

\vspace{3pt}
\noindent\textbf{$\bullet$ Multimodal Content Generation.}
A lot of prior studies utilize pre-trained generative models for content generation across various modalities, including image~\cite{koh2024generating,hu2024instruct}, text~\cite{yang2024glyphcontrol,li2023blip}, video~\cite{jin2024video,liu2024sora}, and audio~\cite{xing2024seeing,lam2024efficient,majumder2024tango}. From these works, we have witnessed the impressive capabilities of the pre-trained LMMs, such as GPT-4~\cite{achiam2023gpt}, LLaVA~\cite{liu2024llava}, LaVIT~\cite{lavit}, and Gemini~\cite{team2023gemini} in instruction-following, multimodal content understanding, and generation. However, despite their success, these models primarily generate content conditioning on text prompts or other given modalities, without incorporating users' personalized information. When directly applied to personalized content generation, these models often exhibit suboptimal performance (\cf the empirical results of pre-trained LaVIT and LLaVA in Table~\ref{table:overall}) due to their limited understanding of user preferences. Therefore, we propose the Pigeon framework, empowering the pre-trained LMMs with personalization capabilities.

\vspace{0.1cm}

\section{Conclusion and Future Work}
\label{sec:conclusion}

In this work, we proposed a novel framework named Pigeon, which integrates a pre-trained LMM with specialized modules to infer implicit user preferences from noisy user history and incorporate explicit multimodal instructions for personalized image generation. To alleviate data scarcity, Pigeon adopts a two-stage preference alignment scheme with masked preference reconstruction and pairwise preference alignment, enhancing the personalization capabilities of LMMs. Both quantitative results and human evaluation highlight Pigeon's effectiveness in generating personalized images. 

This work marks an initial attempt to align pre-trained LMMs with implicit user visual preferences, paving the way for several promising directions: 1) adapting Pigeon to consider evolving user preferences; 2) developing efficient strategies to manage lifelong user history for superior personalization; 3) integrating personalized content generation and filtering to construct more powerful personalized content delivery systems.

\clearpage

{
\tiny
\bibliographystyle{ACM-Reference-Format}
\balance
\bibliography{bibfile}


\begin{thebibliography}{57}


\ifx \showCODEN    \undefined \def \showCODEN     #1{\unskip}     \fi
\ifx \showISBNx    \undefined \def \showISBNx     #1{\unskip}     \fi
\ifx \showISBNxiii \undefined \def \showISBNxiii  #1{\unskip}     \fi
\ifx \showISSN     \undefined \def \showISSN      #1{\unskip}     \fi
\ifx \showLCCN     \undefined \def \showLCCN      #1{\unskip}     \fi
\ifx \shownote     \undefined \def \shownote      #1{#1}          \fi
\ifx \showarticletitle \undefined \def \showarticletitle #1{#1}   \fi
\ifx \showURL      \undefined \def \showURL       {\relax}        \fi
\providecommand\bibfield[2]{#2}
\providecommand\bibinfo[2]{#2}
\providecommand\natexlab[1]{#1}
\providecommand\showeprint[2][]{arXiv:#2}

\bibitem[Achiam et~al\mbox{.}(2023)]%
        {achiam2023gpt}
\bibfield{author}{\bibinfo{person}{Josh Achiam}, \bibinfo{person}{Steven Adler}, \bibinfo{person}{Sandhini Agarwal}, \bibinfo{person}{Lama Ahmad}, \bibinfo{person}{Ilge Akkaya}, \bibinfo{person}{Florencia~Leoni Aleman}, \bibinfo{person}{Diogo Almeida}, \bibinfo{person}{Janko Altenschmidt}, \bibinfo{person}{Sam Altman}, \bibinfo{person}{Shyamal Anadkat}, {et~al\mbox{.}}} \bibinfo{year}{2023}\natexlab{}.
\newblock \showarticletitle{Gpt-4 technical report}.
\newblock \bibinfo{journal}{\emph{arXiv:2303.08774}} (\bibinfo{year}{2023}).
\newblock


\bibitem[Bao et~al\mbox{.}(2023)]%
        {bao2023tallrec}
\bibfield{author}{\bibinfo{person}{Keqin Bao}, \bibinfo{person}{Jizhi Zhang}, \bibinfo{person}{Yang Zhang}, \bibinfo{person}{Wenjie Wang}, \bibinfo{person}{Fuli Feng}, {and} \bibinfo{person}{Xiangnan He}.} \bibinfo{year}{2023}\natexlab{}.
\newblock \showarticletitle{Tallrec: An effective and efficient tuning framework to align large language model with recommendation}. In \bibinfo{booktitle}{\emph{RecSys}}. \bibinfo{publisher}{ACM}, \bibinfo{pages}{1007--1014}.
\newblock


\bibitem[Czapp et~al\mbox{.}(2024)]%
        {czapp2024dynamic}
\bibfield{author}{\bibinfo{person}{\'{A}d\'{a}m~Tibor Czapp}, \bibinfo{person}{M\'{a}ty\'{a}s Jani}, \bibinfo{person}{B\'{a}lint Domi\'{a}n}, {and} \bibinfo{person}{Bal\'{a}zs Hidasi}.} \bibinfo{year}{2024}\natexlab{}.
\newblock \showarticletitle{Dynamic Product Image Generation and Recommendation at Scale for Personalized E-commerce}. In \bibinfo{booktitle}{\emph{RecSys}}. \bibinfo{publisher}{ACM}, \bibinfo{pages}{768–770}.
\newblock


\bibitem[Dai et~al\mbox{.}(2022)]%
        {dai2022personalised}
\bibfield{author}{\bibinfo{person}{Shuqi Dai}, \bibinfo{person}{Xichu Ma}, \bibinfo{person}{Ye Wang}, {and} \bibinfo{person}{Roger~B Dannenberg}.} \bibinfo{year}{2022}\natexlab{}.
\newblock \showarticletitle{Personalised popular music generation using imitation and structure}.
\newblock \bibinfo{journal}{\emph{J New Music Res}} \bibinfo{volume}{51}, \bibinfo{number}{1} (\bibinfo{year}{2022}), \bibinfo{pages}{69--85}.
\newblock


\bibitem[Gal et~al\mbox{.}(2023)]%
        {textualinversion}
\bibfield{author}{\bibinfo{person}{Rinon Gal}, \bibinfo{person}{Yuval Alaluf}, \bibinfo{person}{Yuval Atzmon}, \bibinfo{person}{Or Patashnik}, \bibinfo{person}{Amit~Haim Bermano}, \bibinfo{person}{Gal Chechik}, {and} \bibinfo{person}{Daniel Cohen{-}Or}.} \bibinfo{year}{2023}\natexlab{}.
\newblock \showarticletitle{An Image is Worth One Word: Personalizing Text-to-Image Generation using Textual Inversion}. In \bibinfo{booktitle}{\emph{ICLR}}. \bibinfo{publisher}{OpenReview.net}.
\newblock


\bibitem[Gao et~al\mbox{.}(2023)]%
        {gao2023cirs}
\bibfield{author}{\bibinfo{person}{Chongming Gao}, \bibinfo{person}{Shiqi Wang}, \bibinfo{person}{Shijun Li}, \bibinfo{person}{Jiawei Chen}, \bibinfo{person}{Xiangnan He}, \bibinfo{person}{Wenqiang Lei}, \bibinfo{person}{Biao Li}, \bibinfo{person}{Yuan Zhang}, {and} \bibinfo{person}{Peng Jiang}.} \bibinfo{year}{2023}\natexlab{}.
\newblock \showarticletitle{CIRS: Bursting Filter Bubbles by Counterfactual Interactive Recommender System}.
\newblock \bibinfo{journal}{\emph{TOIS}} \bibinfo{volume}{42}, \bibinfo{number}{1}, Article \bibinfo{articleno}{14} (\bibinfo{year}{2023}).
\newblock


\bibitem[Ge et~al\mbox{.}(2024)]%
        {ge2024making}
\bibfield{author}{\bibinfo{person}{Yuying Ge}, \bibinfo{person}{Sijie Zhao}, \bibinfo{person}{Ziyun Zeng}, \bibinfo{person}{Yixiao Ge}, \bibinfo{person}{Chen Li}, \bibinfo{person}{Xintao Wang}, {and} \bibinfo{person}{Ying Shan}.} \bibinfo{year}{2024}\natexlab{}.
\newblock \showarticletitle{Making {LL}a{MA} {SEE} and Draw with {SEED} Tokenizer}. In \bibinfo{booktitle}{\emph{ICLR}}. \bibinfo{publisher}{OpenReview.net}.
\newblock


\bibitem[He et~al\mbox{.}(2020)]%
        {he2020lightgcn}
\bibfield{author}{\bibinfo{person}{Xiangnan He}, \bibinfo{person}{Kuan Deng}, \bibinfo{person}{Xiang Wang}, \bibinfo{person}{Yan Li}, \bibinfo{person}{Yongdong Zhang}, {and} \bibinfo{person}{Meng Wang}.} \bibinfo{year}{2020}\natexlab{}.
\newblock \showarticletitle{Lightgcn: Simplifying and powering graph convolution network for recommendation}. In \bibinfo{booktitle}{\emph{SIGIR}}. \bibinfo{publisher}{ACM}, \bibinfo{pages}{639--648}.
\newblock


\bibitem[He et~al\mbox{.}(2024)]%
        {he2024imagine}
\bibfield{author}{\bibinfo{person}{Zecheng He}, \bibinfo{person}{Bo Sun}, \bibinfo{person}{Felix Juefei-Xu}, \bibinfo{person}{Haoyu Ma}, \bibinfo{person}{Ankit Ramchandani}, \bibinfo{person}{Vincent Cheung}, \bibinfo{person}{Siddharth Shah}, \bibinfo{person}{Anmol Kalia}, \bibinfo{person}{Harihar Subramanyam}, \bibinfo{person}{Alireza Zareian}, {et~al\mbox{.}}} \bibinfo{year}{2024}\natexlab{}.
\newblock \showarticletitle{Imagine yourself: Tuning-Free Personalized Image Generation}.
\newblock \bibinfo{journal}{\emph{arXiv:2409.13346}} (\bibinfo{year}{2024}).
\newblock


\bibitem[Hu et~al\mbox{.}(2022)]%
        {hu2022lora}
\bibfield{author}{\bibinfo{person}{Edward~J Hu}, \bibinfo{person}{yelong shen}, \bibinfo{person}{Phillip Wallis}, \bibinfo{person}{Zeyuan Allen-Zhu}, \bibinfo{person}{Yuanzhi Li}, \bibinfo{person}{Shean Wang}, \bibinfo{person}{Lu Wang}, {and} \bibinfo{person}{Weizhu Chen}.} \bibinfo{year}{2022}\natexlab{}.
\newblock \showarticletitle{Lo{RA}: Low-Rank Adaptation of Large Language Models}. In \bibinfo{booktitle}{\emph{ICLR}}. \bibinfo{publisher}{OpenReview.net}.
\newblock


\bibitem[Hu et~al\mbox{.}(2024)]%
        {hu2024instruct}
\bibfield{author}{\bibinfo{person}{Hexiang Hu}, \bibinfo{person}{Kelvin~CK Chan}, \bibinfo{person}{Yu-Chuan Su}, \bibinfo{person}{Wenhu Chen}, \bibinfo{person}{Yandong Li}, \bibinfo{person}{Kihyuk Sohn}, \bibinfo{person}{Yang Zhao}, \bibinfo{person}{Xue Ben}, \bibinfo{person}{Boqing Gong}, \bibinfo{person}{William Cohen}, {et~al\mbox{.}}} \bibinfo{year}{2024}\natexlab{}.
\newblock \showarticletitle{Instruct-Imagen: Image generation with multi-modal instruction}. In \bibinfo{booktitle}{\emph{CVPR}}. \bibinfo{publisher}{IEEE}, \bibinfo{pages}{4754--4763}.
\newblock


\bibitem[Jin et~al\mbox{.}(2024a)]%
        {jin2024video}
\bibfield{author}{\bibinfo{person}{Yang Jin}, \bibinfo{person}{Zhicheng Sun}, \bibinfo{person}{Kun Xu}, \bibinfo{person}{Liwei Chen}, \bibinfo{person}{Hao Jiang}, \bibinfo{person}{Quzhe Huang}, \bibinfo{person}{Chengru Song}, \bibinfo{person}{Yuliang Liu}, \bibinfo{person}{Di Zhang}, \bibinfo{person}{Yang Song}, {et~al\mbox{.}}} \bibinfo{year}{2024}\natexlab{a}.
\newblock \showarticletitle{Video-lavit: Unified video-language pre-training with decoupled visual-motional tokenization}.
\newblock \bibinfo{journal}{\emph{arXiv:2402.03161}} (\bibinfo{year}{2024}).
\newblock


\bibitem[Jin et~al\mbox{.}(2024b)]%
        {lavit}
\bibfield{author}{\bibinfo{person}{Yang Jin}, \bibinfo{person}{Kun Xu}, \bibinfo{person}{Kun Xu}, \bibinfo{person}{Liwei Chen}, \bibinfo{person}{Chao Liao}, \bibinfo{person}{Jianchao Tan}, \bibinfo{person}{Quzhe Huang}, \bibinfo{person}{Bin Chen}, \bibinfo{person}{Chengru Song}, \bibinfo{person}{Dai Meng}, \bibinfo{person}{Di Zhang}, \bibinfo{person}{Wenwu Ou}, \bibinfo{person}{Kun Gai}, {and} \bibinfo{person}{Yadong Mu}.} \bibinfo{year}{2024}\natexlab{b}.
\newblock \showarticletitle{Unified Language-Vision Pretraining in {LLM} with Dynamic Discrete Visual Tokenization}. In \bibinfo{booktitle}{\emph{ICLR}}. \bibinfo{publisher}{OpenReview.net}.
\newblock


\bibitem[Koh et~al\mbox{.}(2024)]%
        {koh2024generating}
\bibfield{author}{\bibinfo{person}{Jing~Yu Koh}, \bibinfo{person}{Daniel Fried}, {and} \bibinfo{person}{Russ~R Salakhutdinov}.} \bibinfo{year}{2024}\natexlab{}.
\newblock \showarticletitle{Generating images with multimodal language models}.
\newblock \bibinfo{journal}{\emph{NeurIPS}}  \bibinfo{volume}{36} (\bibinfo{year}{2024}).
\newblock


\bibitem[Kumar et~al\mbox{.}(2024)]%
        {kumar2024longlamp}
\bibfield{author}{\bibinfo{person}{Ishita Kumar}, \bibinfo{person}{Snigdha Viswanathan}, \bibinfo{person}{Sushrita Yerra}, \bibinfo{person}{Alireza Salemi}, \bibinfo{person}{Ryan~A Rossi}, \bibinfo{person}{Franck Dernoncourt}, \bibinfo{person}{Hanieh Deilamsalehy}, \bibinfo{person}{Xiang Chen}, \bibinfo{person}{Ruiyi Zhang}, \bibinfo{person}{Shubham Agarwal}, {et~al\mbox{.}}} \bibinfo{year}{2024}\natexlab{}.
\newblock \showarticletitle{LongLaMP: A Benchmark for Personalized Long-form Text Generation}.
\newblock \bibinfo{journal}{\emph{arXiv:2407.11016}} (\bibinfo{year}{2024}).
\newblock


\bibitem[Kumari et~al\mbox{.}(2023)]%
        {customdiff}
\bibfield{author}{\bibinfo{person}{Nupur Kumari}, \bibinfo{person}{Bingliang Zhang}, \bibinfo{person}{Richard Zhang}, \bibinfo{person}{Eli Shechtman}, {and} \bibinfo{person}{Jun-Yan Zhu}.} \bibinfo{year}{2023}\natexlab{}.
\newblock \showarticletitle{Multi-concept customization of text-to-image diffusion}. In \bibinfo{booktitle}{\emph{CVPR}}. \bibinfo{publisher}{IEEE}, \bibinfo{pages}{1931--1941}.
\newblock


\bibitem[Lam et~al\mbox{.}(2024)]%
        {lam2024efficient}
\bibfield{author}{\bibinfo{person}{Max~WY Lam}, \bibinfo{person}{Qiao Tian}, \bibinfo{person}{Tang Li}, \bibinfo{person}{Zongyu Yin}, \bibinfo{person}{Siyuan Feng}, \bibinfo{person}{Ming Tu}, \bibinfo{person}{Yuliang Ji}, \bibinfo{person}{Rui Xia}, \bibinfo{person}{Mingbo Ma}, \bibinfo{person}{Xuchen Song}, {et~al\mbox{.}}} \bibinfo{year}{2024}\natexlab{}.
\newblock \showarticletitle{Efficient neural music generation}.
\newblock \bibinfo{journal}{\emph{NeurIPS}}  \bibinfo{volume}{36} (\bibinfo{year}{2024}).
\newblock


\bibitem[Li et~al\mbox{.}(2023)]%
        {li2023blip}
\bibfield{author}{\bibinfo{person}{Junnan Li}, \bibinfo{person}{Dongxu Li}, \bibinfo{person}{Silvio Savarese}, {and} \bibinfo{person}{Steven Hoi}.} \bibinfo{year}{2023}\natexlab{}.
\newblock \showarticletitle{Blip-2: Bootstrapping language-image pre-training with frozen image encoders and large language models}. In \bibinfo{booktitle}{\emph{ICML}}. PMLR, \bibinfo{pages}{19730--19742}.
\newblock


\bibitem[Li et~al\mbox{.}(2024)]%
        {li2024stylegan}
\bibfield{author}{\bibinfo{person}{Xiaoming Li}, \bibinfo{person}{Xinyu Hou}, {and} \bibinfo{person}{Chen~Change Loy}.} \bibinfo{year}{2024}\natexlab{}.
\newblock \showarticletitle{When stylegan meets stable diffusion: a w+ adapter for personalized image generation}. In \bibinfo{booktitle}{\emph{CVPR}}. \bibinfo{publisher}{IEEE}, \bibinfo{pages}{2187--2196}.
\newblock


\bibitem[Lin et~al\mbox{.}(2024)]%
        {lin2024bridging}
\bibfield{author}{\bibinfo{person}{Xinyu Lin}, \bibinfo{person}{Wenjie Wang}, \bibinfo{person}{Yongqi Li}, \bibinfo{person}{Fuli Feng}, \bibinfo{person}{See-Kiong Ng}, {and} \bibinfo{person}{Tat-Seng Chua}.} \bibinfo{year}{2024}\natexlab{}.
\newblock \showarticletitle{Bridging Items and Language: A Transition Paradigm for Large Language Model-Based Recommendation}. In \bibinfo{booktitle}{\emph{KDD}}. \bibinfo{publisher}{ACM}, \bibinfo{pages}{1816–1826}.
\newblock


\bibitem[Liu et~al\mbox{.}(2024c)]%
        {ClusterGCF2024TOIS}
\bibfield{author}{\bibinfo{person}{Fan Liu}, \bibinfo{person}{Shuai Zhao}, \bibinfo{person}{Zhiyong Cheng}, \bibinfo{person}{Liqiang Nie}, {and} \bibinfo{person}{Mohan Kankanhalli}.} \bibinfo{year}{2024}\natexlab{c}.
\newblock \showarticletitle{Cluster-Based Graph Collaborative Filtering}.
\newblock \bibinfo{journal}{\emph{TOIS}} \bibinfo{volume}{42}, \bibinfo{number}{6}, Article \bibinfo{articleno}{167} (\bibinfo{year}{2024}).
\newblock


\bibitem[Liu et~al\mbox{.}(2024a)]%
        {liu2024llava}
\bibfield{author}{\bibinfo{person}{Haotian Liu}, \bibinfo{person}{Chunyuan Li}, \bibinfo{person}{Qingyang Wu}, {and} \bibinfo{person}{Yong~Jae Lee}.} \bibinfo{year}{2024}\natexlab{a}.
\newblock \showarticletitle{Visual instruction tuning}.
\newblock \bibinfo{journal}{\emph{NeurIPS}}  \bibinfo{volume}{36} (\bibinfo{year}{2024}).
\newblock


\bibitem[Liu et~al\mbox{.}(2024b)]%
        {liu2024sora}
\bibfield{author}{\bibinfo{person}{Yixin Liu}, \bibinfo{person}{Kai Zhang}, \bibinfo{person}{Yuan Li}, \bibinfo{person}{Zhiling Yan}, \bibinfo{person}{Chujie Gao}, \bibinfo{person}{Ruoxi Chen}, \bibinfo{person}{Zhengqing Yuan}, \bibinfo{person}{Yue Huang}, \bibinfo{person}{Hanchi Sun}, \bibinfo{person}{Jianfeng Gao}, {et~al\mbox{.}}} \bibinfo{year}{2024}\natexlab{b}.
\newblock \showarticletitle{Sora: A review on background, technology, limitations, and opportunities of large vision models}.
\newblock \bibinfo{journal}{\emph{arXiv:2402.17177}} (\bibinfo{year}{2024}).
\newblock


\bibitem[Maddison et~al\mbox{.}(2017)]%
        {gumbel}
\bibfield{author}{\bibinfo{person}{Chris~J. Maddison}, \bibinfo{person}{Andriy Mnih}, {and} \bibinfo{person}{Yee~Whye Teh}.} \bibinfo{year}{2017}\natexlab{}.
\newblock \showarticletitle{The Concrete Distribution: {A} Continuous Relaxation of Discrete Random Variables}. In \bibinfo{booktitle}{\emph{ICLR}}. \bibinfo{publisher}{OpenReview.net}.
\newblock


\bibitem[Majumder et~al\mbox{.}(2024)]%
        {majumder2024tango}
\bibfield{author}{\bibinfo{person}{Navonil Majumder}, \bibinfo{person}{Chia-Yu Hung}, \bibinfo{person}{Deepanway Ghosal}, \bibinfo{person}{Wei-Ning Hsu}, \bibinfo{person}{Rada Mihalcea}, {and} \bibinfo{person}{Soujanya Poria}.} \bibinfo{year}{2024}\natexlab{}.
\newblock \showarticletitle{Tango 2: Aligning Diffusion-based Text-to-Audio Generative Models through Direct Preference Optimization}. In \bibinfo{booktitle}{\emph{MM}}. \bibinfo{publisher}{ACM}.
\newblock


\bibitem[Oquab et~al\mbox{.}(2024)]%
        {oquab2024dinov}
\bibfield{author}{\bibinfo{person}{Maxime Oquab}, \bibinfo{person}{Timoth{\'e}e Darcet}, \bibinfo{person}{Th{\'e}o Moutakanni}, \bibinfo{person}{Huy~V. Vo}, \bibinfo{person}{Marc Szafraniec}, \bibinfo{person}{Vasil Khalidov}, \bibinfo{person}{Pierre Fernandez}, \bibinfo{person}{Daniel HAZIZA}, \bibinfo{person}{Francisco Massa}, \bibinfo{person}{Alaaeldin El-Nouby}, \bibinfo{person}{Mido Assran}, \bibinfo{person}{Nicolas Ballas}, \bibinfo{person}{Wojciech Galuba}, \bibinfo{person}{Russell Howes}, \bibinfo{person}{Po-Yao Huang}, \bibinfo{person}{Shang-Wen Li}, \bibinfo{person}{Ishan Misra}, \bibinfo{person}{Michael Rabbat}, \bibinfo{person}{Vasu Sharma}, \bibinfo{person}{Gabriel Synnaeve}, \bibinfo{person}{Hu Xu}, \bibinfo{person}{Herve Jegou}, \bibinfo{person}{Julien Mairal}, \bibinfo{person}{Patrick Labatut}, \bibinfo{person}{Armand Joulin}, {and} \bibinfo{person}{Piotr Bojanowski}.} \bibinfo{year}{2024}\natexlab{}.
\newblock \showarticletitle{{DINO}v2: Learning Robust Visual Features without Supervision}.
\newblock \bibinfo{journal}{\emph{TMLR}} (\bibinfo{year}{2024}).
\newblock


\bibitem[Plitsis et~al\mbox{.}(2024)]%
        {manos2024music}
\bibfield{author}{\bibinfo{person}{Manos Plitsis}, \bibinfo{person}{Theodoros Kouzelis}, \bibinfo{person}{Georgios Paraskevopoulos}, \bibinfo{person}{Vassilis Katsouros}, {and} \bibinfo{person}{Yannis Panagakis}.} \bibinfo{year}{2024}\natexlab{}.
\newblock \showarticletitle{Investigating Personalization Methods in Text to Music Generation}. In \bibinfo{booktitle}{\emph{ICASSP}}. \bibinfo{publisher}{IEEE}, \bibinfo{pages}{1081--1085}.
\newblock


\bibitem[Podell et~al\mbox{.}(2024)]%
        {podell2024sdxl}
\bibfield{author}{\bibinfo{person}{Dustin Podell}, \bibinfo{person}{Zion English}, \bibinfo{person}{Kyle Lacey}, \bibinfo{person}{Andreas Blattmann}, \bibinfo{person}{Tim Dockhorn}, \bibinfo{person}{Jonas M{\"u}ller}, \bibinfo{person}{Joe Penna}, {and} \bibinfo{person}{Robin Rombach}.} \bibinfo{year}{2024}\natexlab{}.
\newblock \showarticletitle{{SDXL}: Improving Latent Diffusion Models for High-Resolution Image Synthesis}. In \bibinfo{booktitle}{\emph{ICLR}}. \bibinfo{publisher}{OpenReview.net}.
\newblock


\bibitem[Radford et~al\mbox{.}(2021)]%
        {radford2021learning}
\bibfield{author}{\bibinfo{person}{Alec Radford}, \bibinfo{person}{Jong~Wook Kim}, \bibinfo{person}{Chris Hallacy}, \bibinfo{person}{Aditya Ramesh}, \bibinfo{person}{Gabriel Goh}, \bibinfo{person}{Sandhini Agarwal}, \bibinfo{person}{Girish Sastry}, \bibinfo{person}{Amanda Askell}, \bibinfo{person}{Pamela Mishkin}, \bibinfo{person}{Jack Clark}, {et~al\mbox{.}}} \bibinfo{year}{2021}\natexlab{}.
\newblock \showarticletitle{Learning transferable visual models from natural language supervision}. In \bibinfo{booktitle}{\emph{ICML}}. PMLR, \bibinfo{pages}{8748--8763}.
\newblock


\bibitem[Rafailov et~al\mbox{.}(2024)]%
        {rafailov2024direct}
\bibfield{author}{\bibinfo{person}{Rafael Rafailov}, \bibinfo{person}{Archit Sharma}, \bibinfo{person}{Eric Mitchell}, \bibinfo{person}{Christopher~D Manning}, \bibinfo{person}{Stefano Ermon}, {and} \bibinfo{person}{Chelsea Finn}.} \bibinfo{year}{2024}\natexlab{}.
\newblock \showarticletitle{Direct preference optimization: Your language model is secretly a reward model}.
\newblock \bibinfo{journal}{\emph{NeurIPS}}  \bibinfo{volume}{36} (\bibinfo{year}{2024}).
\newblock


\bibitem[Rame et~al\mbox{.}(2024)]%
        {rame2024rewarded}
\bibfield{author}{\bibinfo{person}{Alexandre Rame}, \bibinfo{person}{Guillaume Couairon}, \bibinfo{person}{Corentin Dancette}, \bibinfo{person}{Jean-Baptiste Gaya}, \bibinfo{person}{Mustafa Shukor}, \bibinfo{person}{Laure Soulier}, {and} \bibinfo{person}{Matthieu Cord}.} \bibinfo{year}{2024}\natexlab{}.
\newblock \showarticletitle{Rewarded soups: towards pareto-optimal alignment by interpolating weights fine-tuned on diverse rewards}.
\newblock \bibinfo{journal}{\emph{NeurIPS}}  \bibinfo{volume}{36} (\bibinfo{year}{2024}).
\newblock


\bibitem[Rombach et~al\mbox{.}(2022)]%
        {rombach2022high}
\bibfield{author}{\bibinfo{person}{Robin Rombach}, \bibinfo{person}{Andreas Blattmann}, \bibinfo{person}{Dominik Lorenz}, \bibinfo{person}{Patrick Esser}, {and} \bibinfo{person}{Bj{\"o}rn Ommer}.} \bibinfo{year}{2022}\natexlab{}.
\newblock \showarticletitle{High-resolution image synthesis with latent diffusion models}. In \bibinfo{booktitle}{\emph{CVPR}}. \bibinfo{pages}{10684--10695}.
\newblock


\bibitem[Ruiz et~al\mbox{.}(2023)]%
        {ruiz2023dreambooth}
\bibfield{author}{\bibinfo{person}{Nataniel Ruiz}, \bibinfo{person}{Yuanzhen Li}, \bibinfo{person}{Varun Jampani}, \bibinfo{person}{Yael Pritch}, \bibinfo{person}{Michael Rubinstein}, {and} \bibinfo{person}{Kfir Aberman}.} \bibinfo{year}{2023}\natexlab{}.
\newblock \showarticletitle{Dreambooth: Fine tuning text-to-image diffusion models for subject-driven generation}. In \bibinfo{booktitle}{\emph{CVPR}}. \bibinfo{publisher}{IEEE}, \bibinfo{pages}{22500--22510}.
\newblock


\bibitem[Salemi et~al\mbox{.}(2024a)]%
        {salemi2024optimization}
\bibfield{author}{\bibinfo{person}{Alireza Salemi}, \bibinfo{person}{Surya Kallumadi}, {and} \bibinfo{person}{Hamed Zamani}.} \bibinfo{year}{2024}\natexlab{a}.
\newblock \showarticletitle{Optimization methods for personalizing large language models through retrieval augmentation}. In \bibinfo{booktitle}{\emph{SIGIR}}. \bibinfo{publisher}{ACM}, \bibinfo{pages}{752--762}.
\newblock


\bibitem[Salemi et~al\mbox{.}(2024b)]%
        {salemi-etal-2024-lamp}
\bibfield{author}{\bibinfo{person}{Alireza Salemi}, \bibinfo{person}{Sheshera Mysore}, \bibinfo{person}{Michael Bendersky}, {and} \bibinfo{person}{Hamed Zamani}.} \bibinfo{year}{2024}\natexlab{b}.
\newblock \showarticletitle{{L}a{MP}: When Large Language Models Meet Personalization}. In \bibinfo{booktitle}{\emph{ACL}}. \bibinfo{publisher}{ACL}, \bibinfo{pages}{7370--7392}.
\newblock


\bibitem[Shen et~al\mbox{.}(2024)]%
        {pmg}
\bibfield{author}{\bibinfo{person}{Xiaoteng Shen}, \bibinfo{person}{Rui Zhang}, \bibinfo{person}{Xiaoyan Zhao}, \bibinfo{person}{Jieming Zhu}, {and} \bibinfo{person}{Xi Xiao}.} \bibinfo{year}{2024}\natexlab{}.
\newblock \showarticletitle{PMG: Personalized Multimodal Generation with Large Language Models}. In \bibinfo{booktitle}{\emph{WWW}}. \bibinfo{publisher}{ACM}, \bibinfo{pages}{3833–3843}.
\newblock


\bibitem[Shi et~al\mbox{.}(2024)]%
        {shi2024instantbooth}
\bibfield{author}{\bibinfo{person}{Jing Shi}, \bibinfo{person}{Wei Xiong}, \bibinfo{person}{Zhe Lin}, {and} \bibinfo{person}{Hyun~Joon Jung}.} \bibinfo{year}{2024}\natexlab{}.
\newblock \showarticletitle{Instantbooth: Personalized text-to-image generation without test-time finetuning}. In \bibinfo{booktitle}{\emph{CVPR}}. \bibinfo{publisher}{IEEE}, \bibinfo{pages}{8543--8552}.
\newblock


\bibitem[Shilova et~al\mbox{.}(2023)]%
        {shilova2023adbooster}
\bibfield{author}{\bibinfo{person}{Veronika Shilova}, \bibinfo{person}{Ludovic~Dos Santos}, \bibinfo{person}{Flavian Vasile}, \bibinfo{person}{Ga{\"e}tan Racic}, {and} \bibinfo{person}{Ugo Tanielian}.} \bibinfo{year}{2023}\natexlab{}.
\newblock \showarticletitle{AdBooster: Personalized Ad Creative Generation using Stable Diffusion Outpainting}.
\newblock \bibinfo{journal}{\emph{arXiv:2309.11507}} (\bibinfo{year}{2023}).
\newblock


\bibitem[Tan et~al\mbox{.}(2024)]%
        {tan2024personalized}
\bibfield{author}{\bibinfo{person}{Zhaoxuan Tan}, \bibinfo{person}{Zheyuan Liu}, {and} \bibinfo{person}{Meng Jiang}.} \bibinfo{year}{2024}\natexlab{}.
\newblock \showarticletitle{Personalized Pieces: Efficient Personalized Large Language Models through Collaborative Efforts}.
\newblock \bibinfo{journal}{\emph{arXiv:2406.10471}} (\bibinfo{year}{2024}).
\newblock


\bibitem[Team et~al\mbox{.}(2023)]%
        {team2023gemini}
\bibfield{author}{\bibinfo{person}{Gemini Team}, \bibinfo{person}{Rohan Anil}, \bibinfo{person}{Sebastian Borgeaud}, \bibinfo{person}{Yonghui Wu}, \bibinfo{person}{Jean-Baptiste Alayrac}, \bibinfo{person}{Jiahui Yu}, \bibinfo{person}{Radu Soricut}, \bibinfo{person}{Johan Schalkwyk}, \bibinfo{person}{Andrew~M Dai}, \bibinfo{person}{Anja Hauth}, {et~al\mbox{.}}} \bibinfo{year}{2023}\natexlab{}.
\newblock \showarticletitle{Gemini: a family of highly capable multimodal models}.
\newblock \bibinfo{journal}{\emph{arXiv:2312.11805}} (\bibinfo{year}{2023}).
\newblock


\bibitem[Touvron et~al\mbox{.}(2023)]%
        {touvron2023llama}
\bibfield{author}{\bibinfo{person}{Hugo Touvron}, \bibinfo{person}{Louis Martin}, \bibinfo{person}{Kevin Stone}, \bibinfo{person}{Peter Albert}, \bibinfo{person}{Amjad Almahairi}, \bibinfo{person}{Yasmine Babaei}, \bibinfo{person}{Nikolay Bashlykov}, \bibinfo{person}{Soumya Batra}, \bibinfo{person}{Prajjwal Bhargava}, \bibinfo{person}{Shruti Bhosale}, {et~al\mbox{.}}} \bibinfo{year}{2023}\natexlab{}.
\newblock \showarticletitle{Llama 2: Open foundation and fine-tuned chat models}.
\newblock \bibinfo{journal}{\emph{arXiv:2307.09288}} (\bibinfo{year}{2023}).
\newblock


\bibitem[Vashishtha et~al\mbox{.}(2024)]%
        {vashishtha2024chaining}
\bibfield{author}{\bibinfo{person}{Shanu Vashishtha}, \bibinfo{person}{Abhinav Prakash}, \bibinfo{person}{Lalitesh Morishetti}, \bibinfo{person}{Kaushiki Nag}, \bibinfo{person}{Yokila Arora}, \bibinfo{person}{Sushant Kumar}, {and} \bibinfo{person}{Kannan Achan}.} \bibinfo{year}{2024}\natexlab{}.
\newblock \showarticletitle{Chaining text-to-image and large language model: A novel approach for generating personalized e-commerce banners}. In \bibinfo{booktitle}{\emph{KDD}}. \bibinfo{publisher}{ACM}, \bibinfo{pages}{5825--5835}.
\newblock


\bibitem[Vaswani(2017)]%
        {vaswani2017attention}
\bibfield{author}{\bibinfo{person}{A Vaswani}.} \bibinfo{year}{2017}\natexlab{}.
\newblock \showarticletitle{Attention is all you need}.
\newblock \bibinfo{journal}{\emph{NeurIPS}} (\bibinfo{year}{2017}).
\newblock


\bibitem[Wang et~al\mbox{.}(2023a)]%
        {wang2023generative}
\bibfield{author}{\bibinfo{person}{Wenjie Wang}, \bibinfo{person}{Xinyu Lin}, \bibinfo{person}{Fuli Feng}, \bibinfo{person}{Xiangnan He}, {and} \bibinfo{person}{Tat-Seng Chua}.} \bibinfo{year}{2023}\natexlab{a}.
\newblock \showarticletitle{Generative recommendation: Towards next-generation recommender paradigm}.
\newblock \bibinfo{journal}{\emph{arXiv:2304.03516}} (\bibinfo{year}{2023}).
\newblock


\bibitem[Wang et~al\mbox{.}(2023b)]%
        {wang2023diffusion}
\bibfield{author}{\bibinfo{person}{Wenjie Wang}, \bibinfo{person}{Yiyan Xu}, \bibinfo{person}{Fuli Feng}, \bibinfo{person}{Xinyu Lin}, \bibinfo{person}{Xiangnan He}, {and} \bibinfo{person}{Tat-Seng Chua}.} \bibinfo{year}{2023}\natexlab{b}.
\newblock \showarticletitle{Diffusion recommender model}. In \bibinfo{booktitle}{\emph{SIGIR}}. \bibinfo{publisher}{ACM}, \bibinfo{pages}{832--841}.
\newblock


\bibitem[Wang et~al\mbox{.}(2019)]%
        {wang2019neural}
\bibfield{author}{\bibinfo{person}{Xiang Wang}, \bibinfo{person}{Xiangnan He}, \bibinfo{person}{Meng Wang}, \bibinfo{person}{Fuli Feng}, {and} \bibinfo{person}{Tat-Seng Chua}.} \bibinfo{year}{2019}\natexlab{}.
\newblock \showarticletitle{Neural graph collaborative filtering}. In \bibinfo{booktitle}{\emph{SIGIR}}. \bibinfo{publisher}{ACM}, \bibinfo{pages}{165--174}.
\newblock


\bibitem[Wang et~al\mbox{.}(2003)]%
        {wang2003multiscale}
\bibfield{author}{\bibinfo{person}{Zhou Wang}, \bibinfo{person}{Eero~P Simoncelli}, {and} \bibinfo{person}{Alan~C Bovik}.} \bibinfo{year}{2003}\natexlab{}.
\newblock \showarticletitle{Multiscale structural similarity for image quality assessment}. In \bibinfo{booktitle}{\emph{ACSSC}}. IEEE, \bibinfo{pages}{1398--1402}.
\newblock


\bibitem[Wu et~al\mbox{.}(2023)]%
        {wu2023personalized}
\bibfield{author}{\bibinfo{person}{Chuhan Wu}, \bibinfo{person}{Fangzhao Wu}, \bibinfo{person}{Yongfeng Huang}, {and} \bibinfo{person}{Xing Xie}.} \bibinfo{year}{2023}\natexlab{}.
\newblock \showarticletitle{Personalized news recommendation: Methods and challenges}.
\newblock \bibinfo{journal}{\emph{TOIS}} \bibinfo{volume}{41}, \bibinfo{number}{1} (\bibinfo{year}{2023}), \bibinfo{pages}{1--50}.
\newblock


\bibitem[Wu et~al\mbox{.}(2013)]%
        {wu2013data}
\bibfield{author}{\bibinfo{person}{Xindong Wu}, \bibinfo{person}{Xingquan Zhu}, \bibinfo{person}{Gong-Qing Wu}, {and} \bibinfo{person}{Wei Ding}.} \bibinfo{year}{2013}\natexlab{}.
\newblock \showarticletitle{Data mining with big data}.
\newblock \bibinfo{journal}{\emph{TKDE}} \bibinfo{volume}{26}, \bibinfo{number}{1} (\bibinfo{year}{2013}), \bibinfo{pages}{97--107}.
\newblock


\bibitem[Xing et~al\mbox{.}(2024)]%
        {xing2024seeing}
\bibfield{author}{\bibinfo{person}{Yazhou Xing}, \bibinfo{person}{Yingqing He}, \bibinfo{person}{Zeyue Tian}, \bibinfo{person}{Xintao Wang}, {and} \bibinfo{person}{Qifeng Chen}.} \bibinfo{year}{2024}\natexlab{}.
\newblock \showarticletitle{Seeing and hearing: Open-domain visual-audio generation with diffusion latent aligners}. In \bibinfo{booktitle}{\emph{CVPR}}. \bibinfo{publisher}{IEEE}, \bibinfo{pages}{7151--7161}.
\newblock


\bibitem[Xu et~al\mbox{.}(2024)]%
        {xu2024diffusion}
\bibfield{author}{\bibinfo{person}{Yiyan Xu}, \bibinfo{person}{Wenjie Wang}, \bibinfo{person}{Fuli Feng}, \bibinfo{person}{Yunshan Ma}, \bibinfo{person}{Jizhi Zhang}, {and} \bibinfo{person}{Xiangnan He}.} \bibinfo{year}{2024}\natexlab{}.
\newblock \showarticletitle{Diffusion Models for Generative Outfit Recommendation}. In \bibinfo{booktitle}{\emph{SIGIR}}. \bibinfo{publisher}{ACM}, \bibinfo{pages}{1350--1359}.
\newblock


\bibitem[Yang et~al\mbox{.}(2024b)]%
        {yang2024new}
\bibfield{author}{\bibinfo{person}{Hao Yang}, \bibinfo{person}{Jianxin Yuan}, \bibinfo{person}{Shuai Yang}, \bibinfo{person}{Linhe Xu}, \bibinfo{person}{Shuo Yuan}, {and} \bibinfo{person}{Yifan Zeng}.} \bibinfo{year}{2024}\natexlab{b}.
\newblock \showarticletitle{A New Creative Generation Pipeline for Click-Through Rate with Stable Diffusion Model}. In \bibinfo{booktitle}{\emph{Companion WWW}}. \bibinfo{publisher}{ACM}, \bibinfo{pages}{180--189}.
\newblock


\bibitem[Yang et~al\mbox{.}(2024a)]%
        {yang2024glyphcontrol}
\bibfield{author}{\bibinfo{person}{Yukang Yang}, \bibinfo{person}{Dongnan Gui}, \bibinfo{person}{Yuhui Yuan}, \bibinfo{person}{Weicong Liang}, \bibinfo{person}{Haisong Ding}, \bibinfo{person}{Han Hu}, {and} \bibinfo{person}{Kai Chen}.} \bibinfo{year}{2024}\natexlab{a}.
\newblock \showarticletitle{Glyphcontrol: Glyph conditional control for visual text generation}.
\newblock \bibinfo{journal}{\emph{NeurIPS}}  \bibinfo{volume}{36} (\bibinfo{year}{2024}).
\newblock


\bibitem[Yu et~al\mbox{.}(2019)]%
        {yu2019personalized}
\bibfield{author}{\bibinfo{person}{Cong Yu}, \bibinfo{person}{Yang Hu}, \bibinfo{person}{Yan Chen}, {and} \bibinfo{person}{Bing Zeng}.} \bibinfo{year}{2019}\natexlab{}.
\newblock \showarticletitle{Personalized fashion design}. In \bibinfo{booktitle}{\emph{ICCV}}. \bibinfo{publisher}{IEEE}, \bibinfo{pages}{9046--9055}.
\newblock


\bibitem[Zhang et~al\mbox{.}(2018)]%
        {zhang2018unreasonable}
\bibfield{author}{\bibinfo{person}{Richard Zhang}, \bibinfo{person}{Phillip Isola}, \bibinfo{person}{Alexei~A Efros}, \bibinfo{person}{Eli Shechtman}, {and} \bibinfo{person}{Oliver Wang}.} \bibinfo{year}{2018}\natexlab{}.
\newblock \showarticletitle{The unreasonable effectiveness of deep features as a perceptual metric}. In \bibinfo{booktitle}{\emph{CVPR}}. \bibinfo{publisher}{IEEE}, \bibinfo{pages}{586--595}.
\newblock


\bibitem[Zhao et~al\mbox{.}(2024)]%
        {zhao2024denoising}
\bibfield{author}{\bibinfo{person}{Jujia Zhao}, \bibinfo{person}{Wenjie Wang}, \bibinfo{person}{Yiyan Xu}, \bibinfo{person}{Teng Sun}, \bibinfo{person}{Fuli Feng}, {and} \bibinfo{person}{Tat-Seng Chua}.} \bibinfo{year}{2024}\natexlab{}.
\newblock \showarticletitle{Denoising diffusion recommender model}. In \bibinfo{booktitle}{\emph{SIGIR}}. \bibinfo{publisher}{ACM}, \bibinfo{pages}{1370--1379}.
\newblock


\bibitem[Zhu et~al\mbox{.}(2024)]%
        {zhu2024understanding}
\bibfield{author}{\bibinfo{person}{Zhengbang Zhu}, \bibinfo{person}{Rongjun Qin}, \bibinfo{person}{Junjie Huang}, \bibinfo{person}{Xinyi Dai}, \bibinfo{person}{Yang Yu}, \bibinfo{person}{Yong Yu}, {and} \bibinfo{person}{Weinan Zhang}.} \bibinfo{year}{2024}\natexlab{}.
\newblock \showarticletitle{Understanding or Manipulation: Rethinking Online Performance Gains of Modern Recommender Systems}.
\newblock \bibinfo{journal}{\emph{TOIS}} \bibinfo{volume}{42}, \bibinfo{number}{4} (\bibinfo{year}{2024}), \bibinfo{pages}{1--32}.
\newblock


\end{thebibliography}
}

\appendix
\section{Domain Applications of Pigeon}
\label{sec:application}
Pigeon empowers LMMs with the capability to generate personalized images, which is applicable in various scenarios such as personalized stickers on social media platforms like Twitter and personalized movie posters on platforms like Netflix (see demonstration in Section~\ref{sec:experiment}). Beyond these, we showcase the potential of Pigeon in other representative domains. 

\vspace{3pt}
\noindent\textbf{$\bullet$ E-commerce: personalized product images.} 
In e-commerce, compelling product images are crucial for drawing attention and driving purchase decisions. Pigeon can analyze user visual preferences from their behaviors to generate personalized product images that match individual tastes in personalized display style and background, delivering a more customized shopping experience.

\vspace{3pt}
\noindent\textbf{$\bullet$ Advertising: personalized advertisements.} 
Pigeon can assist advertisers in creating highly customized and context-aware multimodal advertisements based on user behaviors, which are more likely to improve user engagement and conversion rates. 

\vspace{3pt}
\noindent\textbf{$\bullet$ Fashion: personalized fashion designs.} 
Pigeon can infer users’ fashion preferences to generate personalized designs for fashion products like clothing, shoes, and jewelry. 
Besides, both fashion designers and users can provide their preferred fashion images with explicit multimodal instructions for Pigeon to customize designs, fostering an interactive and collaborative design experience. 

\section{Datasets}
\label{sec:appendix_data}
For the sticker scenario, we exclude low-quality themes or those with fewer than six stickers, constructing user interaction sequences where each user interacts with a single theme. For the movie scenario, we adopt the small version of the dataset, retaining user interactions with ratings of four or higher, sorted by the timestamps. We apply a sliding window of six interactions, moving one step at a time to create data samples for each user in both scenarios. Each sample treats the first five interactions as the user history images and the last as the target image. 
We split the samples into training, validation, and testing sets with a ratio of 8:1:1. In the sticker testing set, we randomly select one sticker from a different theme than the user history as the reference image, while in the movie poster scenario, the target image is used as the reference. Dataset statistics are summarized in Table~\ref{tab:dataset}, where each ``sample" consists of user-interacted history images and one reference image.

\begin{table}[H]
\setlength{\abovecaptionskip}{0.05cm}
\setlength{\belowcaptionskip}{0cm}
\caption{Overview of dataset statistics.}
\label{tab:dataset}
\setlength{\abovecaptionskip}{0cm}
\setlength{\belowcaptionskip}{-0.5cm}
\begin{tabular}{cccc}
\hline
\multicolumn{1}{l}{}   & \textbf{\#Users} & \textbf{\#Items} & \textbf{\#Samples} \\ \hline
\textbf{Stickers}      & 725              & 14,345           & 10,719             \\
\textbf{Movie posters} & 594              & 6,961            & 31,058             \\ \hline
\end{tabular}
\end{table}

\section{Implementation Details of Pigeon}
\label{sec:appendix_cost}

In Pigeon, the learning rate is set to $1e^{-5}$ and $5e^{-6}$ for the first and second stage alignment, respectively. The history mask ratio $\alpha_h$ is fixed at $0.2$. During inference, we select the optimal reference mask ratio $\alpha_r\in\{0.0,0.1,\dots,1.0\}$ for each reference image by averaging the history CIS and reference CS.

All experiments are conducted using a single NVIDIA-A100 GPU. As shown in Table~\ref{tab:cost}, while the total number of parameters in Pigeon is substantial, the trainable components represent only a small fraction, leading to relatively low computational overhead. The training process costs about 20 hours and 5 hours for the first and second stage alignment, respectively. For inference, each sample takes about 7 seconds for LaVIT and 9 seconds for SDXL.

\begin{table}[ht]
\setlength{\abovecaptionskip}{0.05cm}
\setlength{\belowcaptionskip}{0cm}
\caption{Model parameters and trainable ratio of Pigeon.}
\label{tab:cost}
\resizebox{0.48\textwidth}{!}{
\begin{tabular}{cl|rc}
\hline
\multicolumn{1}{l}{}                                     &                         & \multicolumn{1}{c}{\textbf{Parameters}} & \textbf{Trainable Ratio} \\ \hline
\multicolumn{2}{c|}{\textbf{Total: LaVIT + SDXL}}                                  & 11,468,249,325                          & -                        \\ \hline
\multicolumn{1}{c|}{\multirow{3}{*}{\textbf{Trainable}}} & \textbf{Mask Generator} & 100,726,784                             & 0.878\%                  \\
\multicolumn{1}{c|}{}                                    & \textbf{Adapter Layer}  & 3,150,336                               & 0.027\%                  \\
\multicolumn{1}{c|}{}                                    & \textbf{LoRA}           & 4,194,304                               & 0.037\%                  \\ \hline
\end{tabular}
}
\end{table}


\section{Human Evaluation}
\label{sec:appendix_human_eval}
In the human evaluation, we conduct binary-choice tests across both sticker and movie scenarios, each consisting of 50 cases. To ensure diversity, the sticker cases involve 49 distinct themes and 6 different emotion labels, including anger, fear, happiness, neutral, sadness, and surprise. On the other hand, the movie cases include 21 different movie genres, such as action, animation, horror, romance, and sci-fi. As shown in Table~\ref{table:human_eval}, we perform a total of 10 binary tests, comparing Pigeon with Grd, TI, and PMG. For each test, 50 participants were recruited for evaluation. The qualitative results validate the effectiveness of Pigeon in personalized image generation across these diverse scenarios and contexts.

\section{Analysis of Preference Reward Strategy} 
\label{sec:appendix_reward}
To evaluate the impact of different preference reward strategies during the second-stage alignment, we conduct additional experiments to compare the strategy outlined in Eq.(\ref{eq:preference_score}) with the approach proposed in~\cite{pmg}. From the results shown in Table~\ref{tab:reward_strategy}, we observe the following key findings: 
\begin{itemize}[leftmargin=*]
    \item The second-stage preference alignment significantly improves personalization, despite a slight decline in semantic alignment. This highlights the effectiveness of DPO in better aligning the generation process with user preferences.
    \item The reward strategy employed by Pigeon achieves superior effectiveness in enhancing personalization, with only a minor and acceptable trade-off in semantic alignment. This indicates that the strategy successfully prioritizes user-specific preferences while maintaining a reasonable degree of semantic consistency.
\end{itemize}

\begin{table*}[t]
\setlength{\abovecaptionskip}{0.05cm}
\setlength{\belowcaptionskip}{0cm}
\caption{Effect of different preference reward strategies during the second-stage alignment.}
\label{tab:reward_strategy}
\begin{tabular}{l|ccccc|ccc}
\hline
\multicolumn{1}{c|}{\textbf{\#Sticker}} & \multicolumn{5}{c|}{\textbf{Personalization}}                                          & \multicolumn{3}{c}{\textbf{Semantic Alignment}} \\
\textbf{Pigeon}                         & \textbf{CS$\uparrow$}   & \textbf{CIS$\uparrow$}  & \textbf{DIS$\uparrow$}  & \textbf{LPIPS$\downarrow$} & \textbf{MS-SSIM$\uparrow$} & \textbf{CS$\uparrow$}  & \textbf{CIS$\uparrow$}  & \textbf{DIS$\uparrow$}  \\ \hline
\textbf{Stage-1}                        & 22.03          & 61.64          & 57.26          & \textbf{0.6800} & {\myul 0.1467}      & {\textbf{25.74}}   & {\textbf{50.66}}    & {\textbf{48.34}}    \\
\textbf{- Stage-2~\cite{pmg}}                      & \textbf{24.15} & {\myul 64.85}    & {\myul 59.43}    & {\myul 0.6803}    & 0.1398            & {\myul 25.57}         & {\myul 50.16}          & {\myul 47.92}          \\
\cellcolor[HTML]{EAEAEA}\textbf{- Stage-2-ours}                 & \cellcolor[HTML]{EAEAEA}{\myul 23.69}    & \cellcolor[HTML]{EAEAEA}\textbf{67.65} & \cellcolor[HTML]{EAEAEA}\textbf{62.23} & \cellcolor[HTML]{EAEAEA}0.6814          & \cellcolor[HTML]{EAEAEA}\textbf{0.1568}   & \cellcolor[HTML]{EAEAEA}21.10         & \cellcolor[HTML]{EAEAEA}47.44          & \cellcolor[HTML]{EAEAEA}45.44          \\ \hline
\end{tabular}

\begin{tabular}{l|ccccc|ccc}
\hline
\multicolumn{1}{c|}{\textbf{\#Movie}} & \multicolumn{5}{c|}{\textbf{Personalization}}                                         & \multicolumn{3}{c}{\textbf{Semantic Alignment}}  \\
\textbf{Pigeon}                       & \textbf{CS$\uparrow$}   & \textbf{CIS$\uparrow$}  & \textbf{DIS$\uparrow$} & \textbf{LPIPS$\downarrow$} & \textbf{MS-SSIM$\uparrow$} & \textbf{CS$\uparrow$}   & \textbf{CIS$\uparrow$}  & \textbf{DIS$\uparrow$}  \\ \hline
\textbf{Stage-1}                      & 15.19          & 37.42          & 17.70         & \textbf{0.7496} & \textbf{0.0493}   & {\myul 27.30}    & 47.79          & 41.35          \\
\textbf{- Stage-2~\cite{pmg}}                    & {\myul 15.28}    & {\myul 38.67}    & 19.03         & 0.7509          & 0.0454            & \textbf{27.80} & {\myul 49.34}    & {\myul 43.03}    \\
\cellcolor[HTML]{EAEAEA}\textbf{- Stage-2-ours}               & \cellcolor[HTML]{EAEAEA}\textbf{15.41} & \cellcolor[HTML]{EAEAEA}\textbf{40.16} & \cellcolor[HTML]{EAEAEA}{\myul 21.29}   & \cellcolor[HTML]{EAEAEA}{\myul 0.7508}    & \cellcolor[HTML]{EAEAEA}{\myul 0.0464}      & \cellcolor[HTML]{EAEAEA}26.45          & \cellcolor[HTML]{EAEAEA}\textbf{49.66} & \cellcolor[HTML]{EAEAEA}\textbf{44.07} \\ \hline
\end{tabular}
\end{table*}

\section{Case Study}
\label{sec:appendix_case}
We present five additional Pigeon-generated examples for both sticker and movie poster scenarios, respectively, along with several user-interacted history images and one reference image. 



\begin{figure}[H]
\setlength{\abovecaptionskip}{0.1cm}
\setlength{\belowcaptionskip}{0cm}
\centering
\includegraphics[scale=0.5]{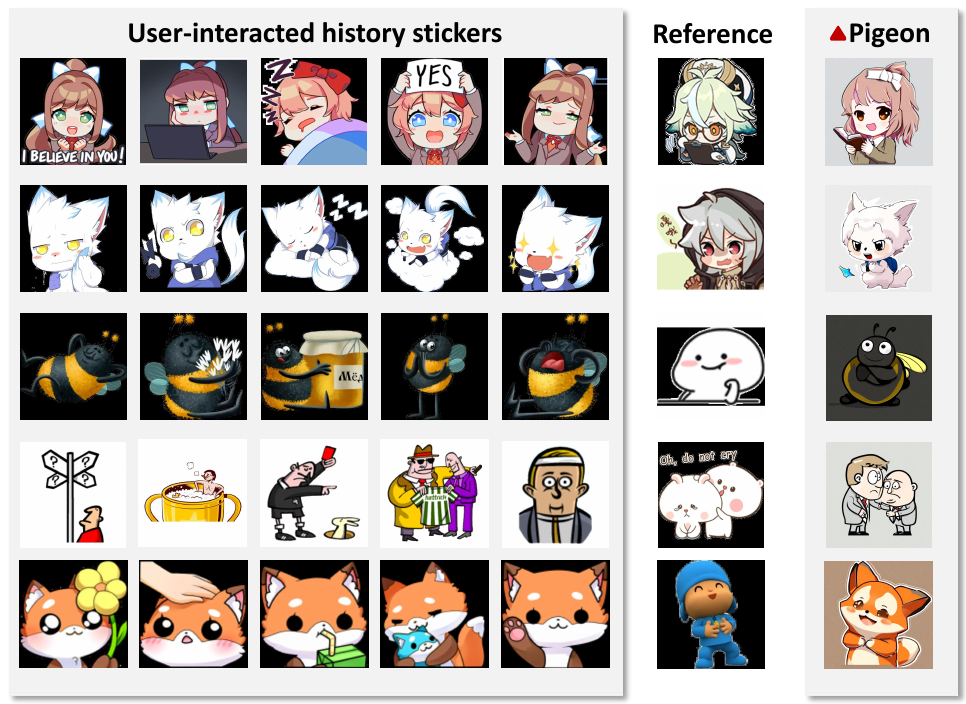}
\caption{Examples of generated stickers, along with user-interacted history stickers and one reference sticker.}
\label{fig:sticker_cases}
\end{figure}

\textbf{Sticker scenario.} As illustrated in Figure~\ref{fig:sticker_cases}, Pigeon effectively captures users' visual preferences for character figures and styles in stickers, and combines these preferences with the high-level semantics of the reference sticker to generate personalized stickers. The generated stickers exhibit high semantic alignment with the reference image, including the conveyed emotions, facial expressions, character actions, and elements like hearts.

\begin{figure}[H]
\setlength{\abovecaptionskip}{0.1cm}
\setlength{\belowcaptionskip}{-0.5cm}
\centering
\includegraphics[scale=0.49]{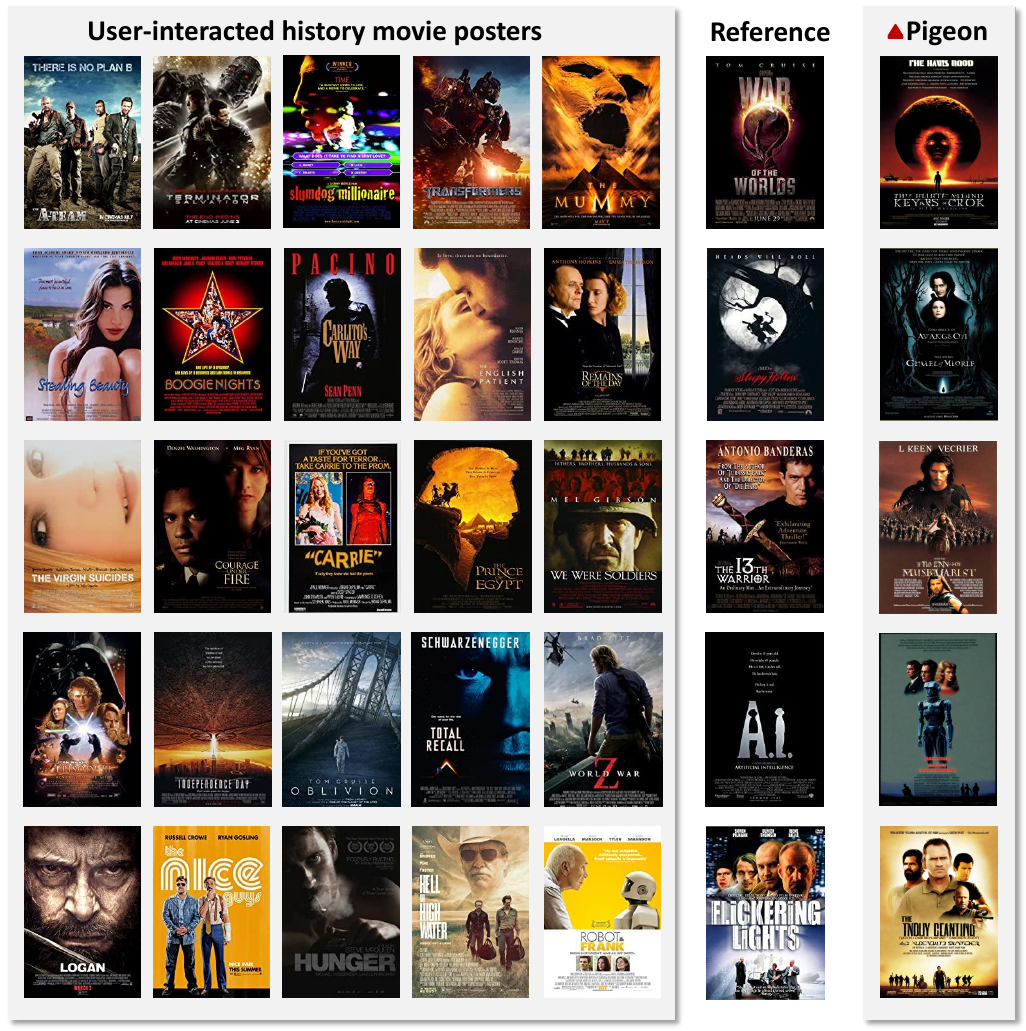}
\caption{Examples of generated movie posters, along with user-interacted history posters and one reference poster.}
\label{fig:movie_cases}
\end{figure}

\textbf{Movie poster scenario.} As depicted in Figure~\ref{fig:movie_cases}, each user shows a distinct set of visual preferences, ranging from action and sci-fi to historical drama and crime thrillers. Pigeon-generated posters effectively mirror these preferences through character-centered designs, dynamic compositions, and color palettes that align with each user's unique taste. By tailoring its designs to the emotional tone, genre, and thematic focus of the reference posters, Pigeon creates personalized posters that strongly resonate with individual users' past interactions and preferences. For instance, the user in the first row shows a strong preference for action-heavy, explosive films with a focus on dramatic visuals and blockbuster-style presentations. Pigeon matches the user’s love for explosive visuals, with characters taking center stage and environments filled with dynamic elements like fire, destruction, and warfare.

\end{document}